\documentclass[%
 reprint,
 amsmath,amssymb,
 aps,pre
]{revtex4-1}
\usepackage{dcolumn}
\usepackage{amssymb}
\usepackage{amsmath}
\usepackage{amsfonts}
\usepackage{amsbsy}
\usepackage{amscd}
\usepackage{amsthm}
\usepackage{times}
\usepackage{psfrag}
\usepackage{graphics}
\usepackage{color}
\usepackage{bbm}
\usepackage{colortbl}
\usepackage{graphicx}
\usepackage{subfigure}
\usepackage{caption}
\usepackage{bbm}
\usepackage{array,supertabular}
\usepackage{tabularx}
\usepackage{booktabs}
\usepackage{multirow}
\usepackage{ulem}
\usepackage{units}
\usepackage{accents}
\usepackage{algorithmicx}
\usepackage{relsize}
\usepackage{algorithm}
\usepackage{xfrac}
\usepackage{tikz}
\usepackage{tikz-3dplot}
\usetikzlibrary{shadings}
\usetikzlibrary{arrows.meta} 
\usetikzlibrary{shapes.misc}
\usetikzlibrary{decorations.pathmorphing,patterns}
\usetikzlibrary{calc,patterns,decorations.markings}
\usepackage{algpseudocode}
\usepackage{rotating}
\usepackage[capitalize]{cleveref}
\usepackage{mathrsfs} 
\usepackage{subfig}

\newcommand{\x}{\mathbf{x}}

\newcommand{\y}{\mathbf{y}}

\crefname{secinapp}{Section}{Sections}
\Crefname{secinapp}{Section}{Sections}

\newlength{\dhatheight}

\begin{document}

\title{A Three-Dimensional Mathematical Model of Collagen Contraction}

\author{E. J. Evans}
\email[]{ejevans@math.byu.edu}
\author{J. C. Dallon}
\email[]{dallon@math.byu.edu}
\affiliation{Department of Mathematics,
  Brigham Young University,
  Provo, Utah 84602, USA}
\date{\today}

\begin{abstract}
In this paper, we introduce a three-dimensional mathematical model of collagen contraction with microbuckling based on the two-dimensional model in \cite{Dallon:2014:MMC}.  The model both qualitatively and quantitatively replicates experimental data including lattice contraction over a time course of 40 hours for lattices with various cell densities, cell density profiles within contracted lattices, radial cut angles in lattices, and cell force propagation within a lattice. The importance of the model lattice formation and the crucial nature of its connectivity are discussed including differences with models which do not include microbuckling.  The model suggests that most cells within contracting lattices are engaged in directed motion.
\end{abstract}
\pacs{}
\keywords{Collagen contraction}
\maketitle
\section{Introduction}
Fibroblast populated collagen lattices have been widely studied since first introduced by Bell et.~al.~\cite{Bell:1979:PTS} with the aim of better understanding cell extracellular matrix interactions and wound contraction.  The contraction of the lattice is an irreversible cell mediated process.  In the first 48 hours, remodeling of the matrix is mainly mechanical and there is little cell proliferation \cite{Ehrlich:2006:EMW,Greco:1992:DCD}, extracellular degredation (although matrix metalloproteinases appear to play a role in contraction via cell dynamics) \cite{Martin:2011:EMI}, or extracellular matrix production \cite{Redden:2003:CCC}. There are three proposed mechanisms responsible for the contraction of the collagen lattice in the first 24-48 hours: cell elongation, tractional forces due to cell locomotion, and cell contraction \cite{Dallon:2008:RFC}.

Previous models related to our work can be divided into two categories.  The first category is models of collagen and fibrous structures \cite{Wyart:2008:EFS,Sharma:2016:SCG}.  These models use discrete formations of the the fibrous structure and are most closely related to the work here. In Wyart et.~al. the authors, while not modeling collagen lattices, use two-dimensional random networks of springs to model fibrous networks and numerically derive material properties of the models.  One of the model parameters which they investigate, which is pertinent to our study, is the average coordination number.  It is defined to be $z=\frac{2N_c}{N}$  where $N_c$ is the number of bonds connecting nodes and $N$ is the number of nodes.  A system is isostatic or rigid when $z=2d$ where $d$ is the dimension.  They found that systems where $z<2d$ have a stress-strain relationship with a zero plateau and then a strain-stiffening region.  This is commonly found in fibrous networks including collagen lattices \cite{Licup:2015:SCM,Sharma:2016:SCG}.  More recently, Sharma et. al. \cite{Sharma:2016:SCG} both modeled and measured material properties of collagen gels.  In their model, they use a lattice of nodes connected by elastic elements which have a Young's modulus and a bending modulus.  The elastic energy of the system is minimized when the system is deformed.  The model discussed here does not have a bending modulus; the fibers are elastic ropes or include microbuckling, and cell interactions are included.

The second category is models of cells interacting with fibrous tissue.  These models include models focused on alignment \cite{Barocas:1997:ABT,Dallon:1998:CAC,Olsen:1999:MMA,Dallon:1999:MME,Schluter:2012:CMS,Reinhardt:2014:AMT,Notbohm:2015:MFP}  and those concerned with contraction \cite{Simon:2012:MRB,Simon:2014:CMM,Dallon:2014:MMC}.  Of these, some are force based \cite{Barocas:1997:ABT,Simon:2012:MRB,Simon:2014:CMM,Reinhardt:2014:AMT,Dallon:2014:MMC,Notbohm:2015:MFP}, some are continuum descriptions \cite{Barocas:1997:ABT,Dallon:1998:CAC,Olsen:1999:MMA,Dallon:1999:MME,Schluter:2012:CMS,Simon:2012:MRB,Simon:2014:CMM}, and some more closely resemble our model by treating the fibrous tissue as a discrete structure and focus on forces \cite{Reinhardt:2014:AMT,Notbohm:2015:MFP}.  In \cite{Reinhardt:2014:AMT}, the matrix is modeled with springs and torsional forces, whereas in \cite{Notbohm:2015:MFP} the matrix is modeled with elastic ropes which they call fibers with microbuckling.  The model here is more similar to that in \cite{Notbohm:2015:MFP} but the focus is on lattice contraction and a larger space scale than either of the other two models.

\section{Mathematical model}\label{sec:notation}
The model is a three-dimensional extension of one of the two-dimensional models presented in \cite{Dallon:2014:MMC}.
\subsection{Model of cell--motion}
\begin{figure}[ht]
   \centerline{
   \includegraphics*[width=.8\linewidth]{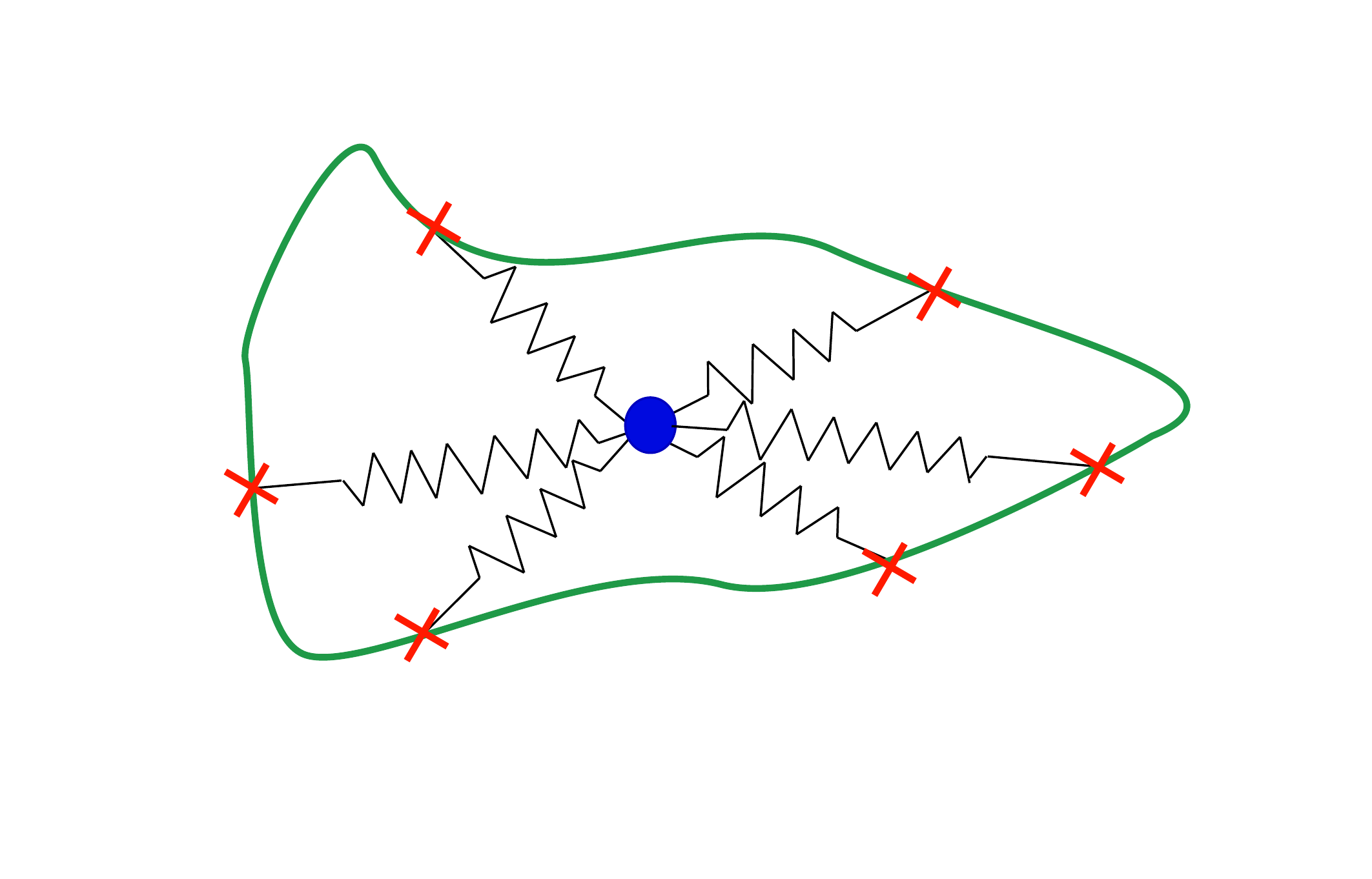}
   } \caption{This figure illustrates how the cell is modeled
   mathematically in this paper.  We consider a cell as a center of mass with attached
   springs.  The other end of the springs are attached to I-sites which
   can interact with the extracellular matrix (membrane bound
   integrin based adhesion sites).  In the simulations for this paper $K=50$; that is, there are 50 I-sites per cell.  Although this figure is presented in the plane, the actual simulations occur in a three-dimensional environment.}
\label{fig:cell}
\end{figure}

The equation of motion for the center of mass of the
$i$th cell with $K$ I-sites is given by
\begin{equation}
C {\bf x}_i^{\prime} = -\sum_{j=1} ^K\alpha(||{\bf x}_i-{\bf y}_{p_{i,j}}||-\ell)\frac{{\bf x}_i-{\bf y}_{p_{i,j}}}{||{\bf x}_i-{\bf y}_{p_{i,j}}||},
\end{equation}
where ${\bf x}_i^{\prime}$ represents the velocity of the cell, $C$ is the drag coefficient, and ${\bf x}_i$ is the location in $\mathbb{R}^3$ of the cell
center for $i=1,\cdots,N$.   The spring constant $\alpha$ is the same for all cells and all I-sites and, together with the spring rest length $\ell$, define the forces exerted by the cell.  In this paper we will fix the number of I-sites per cell $K$
 at 50 and $\ell=0$. The Reynolds number is low and therefore, because of the relative
magnitudes of the coefficients, the expected acceleration term on the right hand side of the equation is set to zero.  For an illustration of how the cell is modeled mathematically see Figure~\ref{fig:cell}.

The I-sites are constrained to attach to lattice nodes, which are specific locations in $\mathbb{R}^3$, that change with time (see Section~\ref{sec:lat}).  The lattice node locations are denoted by ${\bf  y}_k$, and I-sites from the same
cell or other cells can be attached to the same node location. The set of
indices $p_{i,1},p_{i,2}, \cdots,p_{i,K}$ specify the lattice nodes associated with the I-sites of cell $i$. 
If an I-site maintains its connection to a lattice node indefinitely, the cell reaches an equilibrium position and the velocity of the cell is zero.  We therefore require the I-sites to detach from the collagen lattice and reattach.  The duration of attachment is taken from a Poisson-like distribution with mean attachment of 60 seconds.  Upon detaching the I-site immediately reattaches to a node in the lattice;  thus, for all times in our simulation, $K$ attachment sites are maintained.

In a change from earlier work \cite{Dallon:2014:MMC}, the determination of the location of the next I-site is dependent on the direction the cell is moving.  Specifically the I-site is placed in a cone with vertex angle 4 degrees, in the direction of motion.  The exact distance from the center of mass is determined from a uniform distribution between $0$ and $115.726$.  The placement of the I-site is discussed in further detail in Section~\ref{sec:isiteloc}.  The I-site then attaches to the closest lattice node.  For more
information about the I-sites, the reader is referred to a related model
discussed in \cite{Dallon:2013:FBM}.

\subsection{Collagen lattice}\label{sec:lat}
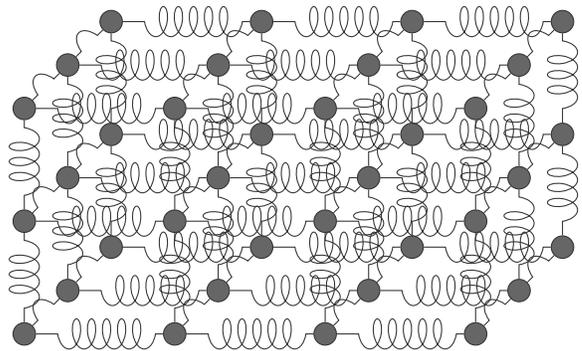
\begin{figure}[h]
\centering
\begin{tikzpicture}[scale=.5,
    decoration={
        coil,
        amplitude = 2mm,
        aspect=.5,
        pre length=1mm,
        post length=1mm,
       segment length=2mm,
    },
    every node/.style={circle, draw}
]

\foreach \x in {0,4,...,12}{
    \let\lasty\undefined
    \foreach \y in {0,3,...,6}{
        \let\lastz\undefined
        \foreach \z in {0,3,6}{
            \begin{scope}[draw=black!60!gray]
                \node[fill=black!60] (n\x\y\z) at (\x,\y,\z) {};
                \ifx\lastx\undefined\else
                    \draw[decorate] (n\x\y\z) -- (n\lastx\y\z);
                \fi
                \ifx\lasty\undefined\else
                    \draw[decorate] (n\x\y\z) -- (n\x\lasty\z);
                \fi
                \ifx\lastz\undefined\else
                    \draw[decorate] (n\x\y\z) -- (n\x\y\lastz);
                \fi
            \end{scope}
            \xdef\lastz{\z}
        };
        \xdef\lasty{\y}
    };
    \xdef\lastx{\x}
};
\end{tikzpicture}
 \caption{This figure illustrates how the collagen lattice is modeled mathematically.  The lattice is defined by nodes with spring-like connections between them. Although this illustration depicts the nodes in a regularly spaced grid and connected to nearest neighbors, the collagen lattices used in our simulation have nodes which are randomly placed and the connections are not restricted to nearest neighbors. }
\label{fig:collagen}
\end{figure}

The collagen lattice is modeled by nodes which are connected with elastic ropes
to form a network of spring-like connections (see Figure~\ref{fig:collagen}).  In \cite{Notbohm:2015:MFP} they call it microbuckling.
To create the collagen lattice, first $M$ nodes are placed in a prescribed domain and a minimal connectivity value $\mathcal{M}$ is specified.
Then, for each node in the lattice the closest $1.5 \times \mathcal{M}$ nodes are selected.  From this selection of nodes exactly $\mathcal{M}$ nodes are sampled without replacement and are connected to the node.  Recall the cell
I-sites are constrained to be at lattice nodes.  

The equation of motion for the lattice node $k$ is
\begin{equation}
\gamma {\bf y}^{\prime}_k(t)=\overbrace{\sum_{m=1,m\neq k}^M {\bf
    f}_{k,m}(t)}^{\mbox{force due to lattice entanglement}}+\overbrace{\sum_{i=1}^N{\bf
    c}_{i,k}(t).}^{\mbox{force due to cells}}
\label{equ:lattice}
\end{equation}
Observe that there are two types of forces acting on lattice nodes.  The first (given by the first summation on the right hand side) is force due to connections with other nodes in the lattice.  The second type of force (given by the second summation on the right hand side) is force due to interactions with cells.  Note that a cell only exerts force directly on a lattice node if the cell has an I-site that is attached to the node.

Forces exerted by other nodes in the lattice can be classified into two types, those forces resulting from normal links and those resulting from compacted links.  
Forces exerted from normal links are spring-like in that if the connection is stretched, the force acts in proportion to the stretching.  If the connection is compressed however, no force is exerted by a normal link.  Compacted links allow for the compaction of collagen and are the result of a non-reversible process.  These links differ from normal links in two key ways.  First, the spring constant is much stiffer than the spring constant for normal links, making the forces exerted by compacted links much higher than the forces exerted by normal links with equivalent amounts of stretching. Second, compacted links resist compression.  The existence of these two types of links are due to the nature of collagen.  When the
collagen fibrils are pulled, they resist the pulling due to their
association with other fibrils. Yet if a cell exerts
forces at two points along the same fibril drawing the two points
closer, the fibril is not compressed but becomes
slack between the two points similar to a rope, i.e., it exhibits microbuckling. 

With these two types of links defined, the force due to a lattice connection between node $k$ and node
$m$ is defined as:
\begin{equation}
{\bf f}_{k,m}(t) = \left \{ 
\begin{array}{ll}
0&\mbox{${\bf y}_k$ and ${\bf y}_m$ are not linked,}\\
0&||{\bf v}_{k,m}||<\ell_{k,m} \,\,\parbox[t][][t]{.2\linewidth}{ and the link\\[-2pt] is normal,}
\\[10pt]
-d_{k,m}{\bf \hat{v}}_{k,m}&
\ell_{k,m}
\leq ||{\bf v}_{k,m}||\,\,\parbox[t][][t]{.2\linewidth}{ and the link\\[-2pt] is normal,}\\[10pt]
-d_{k,m}^*{\bf \hat{v}}_{k,m}&
\mbox{if the
  link is compacted.}
\end{array} \right .
\end{equation}
Here ${\bf y}_m$ is the location of lattice node $m$, ${\bf v}_{k,m}={\bf y}_k-{\bf y}_m$, ${\bf \hat{v}}_{k,m}={\bf v}_{k,m}/||{\bf v}_{k,m}||$, $d_{k,m}=\beta(||{\bf v}_{k,m}||-\ell_{k,m})$ the signed magnitude of the force generated by a normal link between node $k$ and node $m$, $d_{k,m}^*=\beta^*(||{\bf v}_{k,m}||-\ell_{k,m}^*)$ the signed magnitude of the force generated by a compacted link between node $k$ and node $m$, 
$\ell_{k,m}$ is the rest length
of the connection between node $k$ and node $m$ and is set as the
initial distance between the nodes at the beginning of the simulation,
$\beta$ is the spring constant for normal links, $\beta^*=d_{\beta}\beta$ is the
spring constant for compacted links, and
$\ell_{k,m}^*=d_{\ell}\ell_{k,m}$ is the rest length for the spring
connecting node $k$ with node $m$ when the link is compacted.
Initially all links are normal and become compacted if the distance
between two linked nodes becomes small enough, that is, if $d_{k,m} <
d_{p}\ell_{k,m}$.  When links are compacted the rest
length of the spring is shortened ($d_{\ell}<1$), the spring constant
is increased ($d_{\beta}>1$),
and the link resists compression.

The forces exerted on lattice nodes due to the cell $i$ are defined by:
\begin{equation}
{\bf c}_{i,k}(t) = \sum_{j=1}^K \alpha(||{\bf x}_i-{\bf y}_{p_{i,j}}||-\ell)\frac{{\bf x}_i-{\bf y}_{p_{i,j}}}{||{\bf x}_i-{\bf y}_{p_{i,j}}||}\delta(p_{i,j}-k),
\end{equation}
where $\bf{x}_i$ is the cell center location, $\bf{y}_{p_{i,j}}$ is a lattice node location, $\alpha$ is the spring constant and $\ell$ is the rest length
of the integrin.  Here $\delta(0)=1$ and $\delta(x)=0$ for any non-zero $x$ and
indicates whether the $j$th I-site of cell $i$ is interacting with
node $k$.   

\section{Results}
\begin{figure}
\centerline{\includegraphics[width= .6\linewidth]{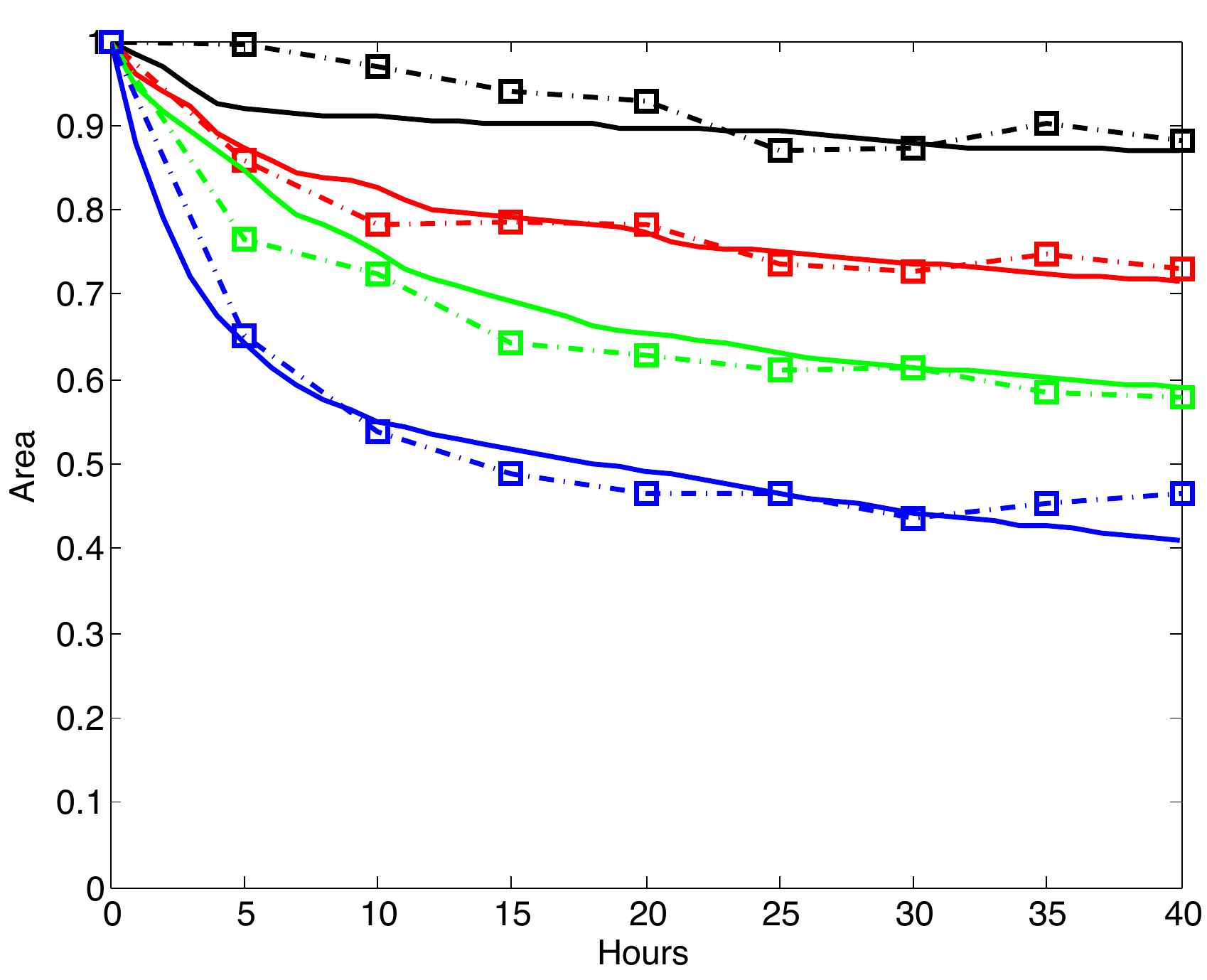}}
\caption{The lattice contraction of four different collagen lattice simulations are
compared to the experimental data detailed in~\cite{Dallon:2014:MMC}.  The squares indicate the experimental values,
and the solid line indicates the results of our simulations. The parameter values for the simulations are 30 for connectivity, 
 $\frac{C}{\beta}= 0.193$ hours,
 $\frac{\alpha}{\beta} = 0.07$, $\frac{\gamma}{\beta}=0.0863$ hours,
 $d_{\beta}=250.538$,
 $d_{\ell}=0.365$, and $d_p=0.365$ microns.  The lines from top to bottom are for cell density 3,750, 10,000, 50,000, and 100,000 cells per mL respectively.  }
\label{fig:comparedata}
\end{figure}

It became clear when extending the model to three dimensions that the lattice formulation and connectivity are crucial to the model results.  We kept the collagen parameters essentially the same as the two-dimensional model.  Only two parameters differ, one is a property of collagen and the other is a property of the cell.  The collagen property that differs is the viscous drag on the collagen nodes, $\frac{\gamma}{\beta}$.  The other parameter is the cell force $\frac{\alpha}{\beta}$.  The collagen property $\frac{\gamma}{\beta}$ used here is 0.0863 compared to 0.114 in the two-dimensional model and for the cell strength $\frac{\alpha}{\beta}$ the value used here is 0.07 and the value used in the two-dimensional model was 2.239.  For the other parameters see the caption for Figure~\ref{fig:comparedata}.

The first objective of our work is to match the experimental observations detailed in~\cite{Dallon:2014:MMC}.  
The results 
for lattices with 3,750, 10,000, 30,000,
and 100,000 cells per mL, gathered over a period of 
40 hours, are shown in Figure~\ref{fig:comparedata}.   
For the numerical simulations, we assume that only 
fibroblasts exist in the collagen lattice.  Note, we do not assume stress dependent attachment mechanisms and stress dependent contraction mechanisms detailed in~\cite{Dallon:2014:MMC}.

\subsection{Lattice connectivity results}
\label{sec:CLR}
When creating a three-dimensional collagen lattice, the three-dimensional analog of the Delaunay triangulation did not create suitable connectivity between nodes in the collagen lattice.  Specifically, there was no guarantee that the nodes would be connected to nodes that were close neighbors.  For this reason we developed our own simple algorithm that guarantees that a node is connected to a minimum of $\mathcal{M}$ neighbors.  There are two key values in our algorithm: first the minimal number of neighbors each node is connected to and second exactly to which neighbors the node is connected.  We address these individually.

\begin{figure}
\centerline{\includegraphics*[width= .5\linewidth]{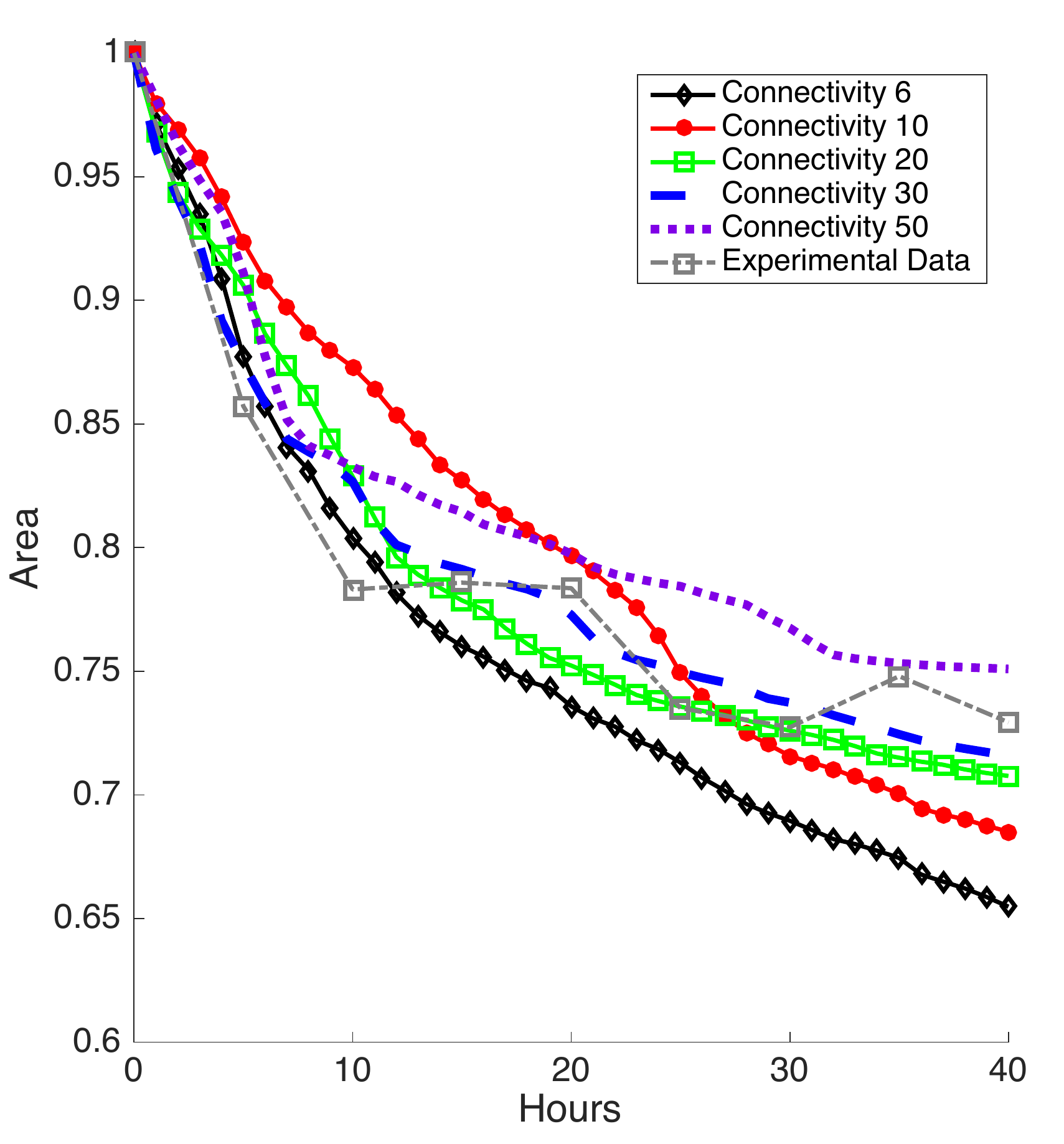}\includegraphics*[width= .5\linewidth]{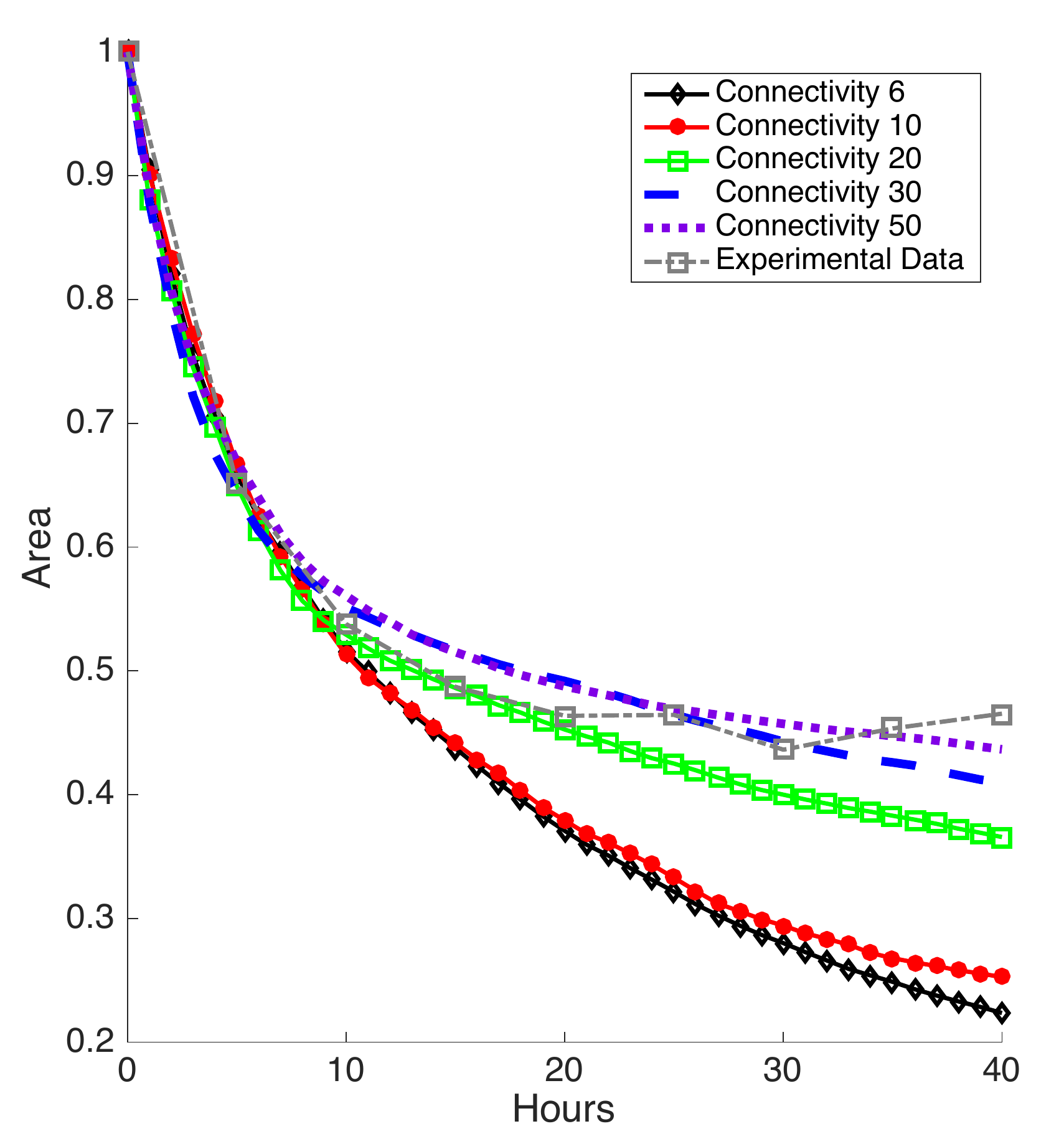}}
\caption{A comparison of contraction for cell densities of 10,000 cells per mL on the left and 100,000 cells per mL on the right. Each line represents a simulation where the lattice has a different connectivity.  Observe that for 10,000 cells per mL the lattice does not contract enough for higher connectivity values, thus indicating that the collagen lattice is too stiff.  For 100,000 cells per mL the lower connectivity values result in a mesh that is too loose and too much collagen contraction occurs.  The squares denote experimental data. }
\label{fig:colcomp}
\end{figure}

\begin{figure}
\centerline{\includegraphics*[width= .5\linewidth]{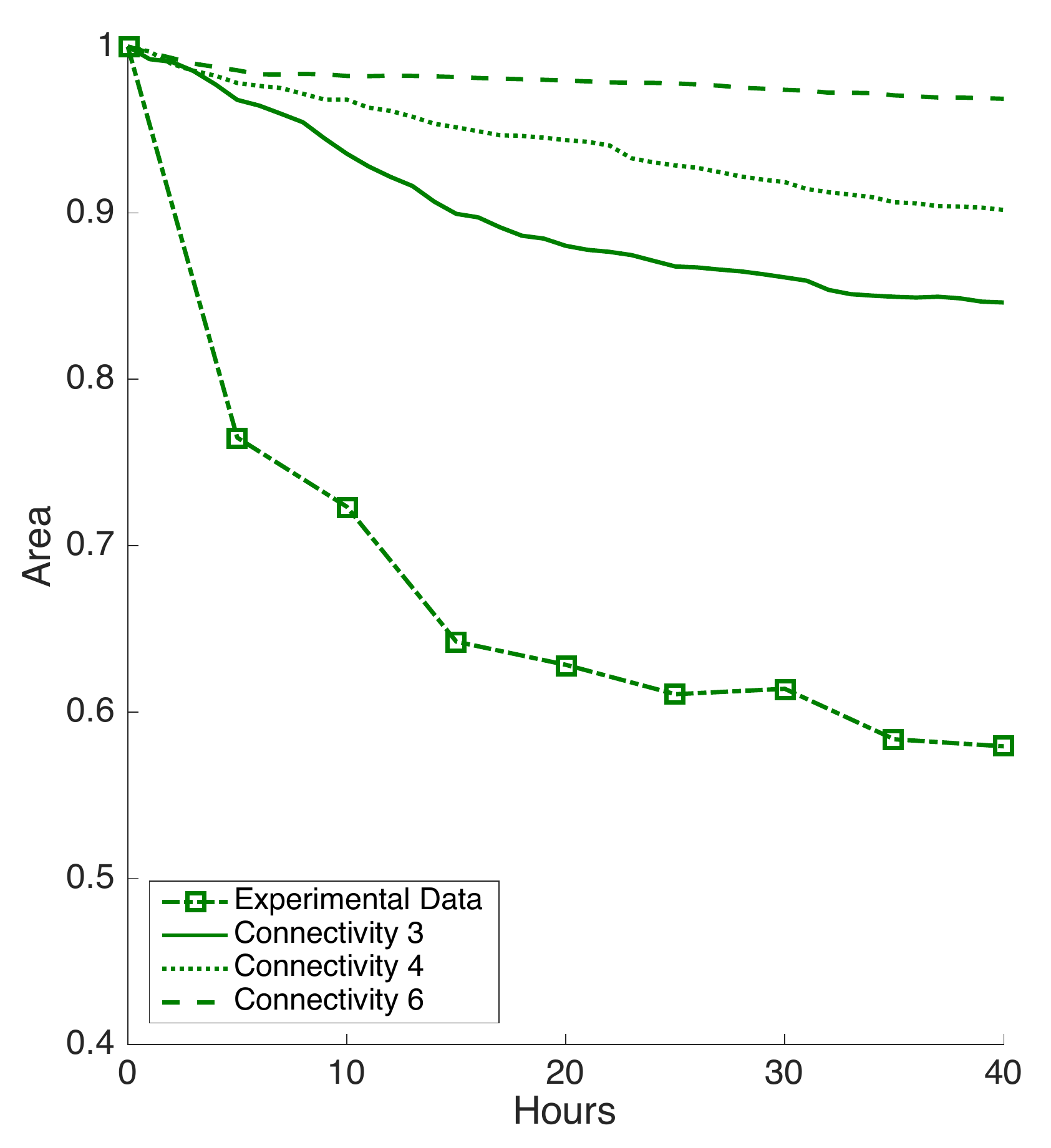}\includegraphics*[width= .5\linewidth]{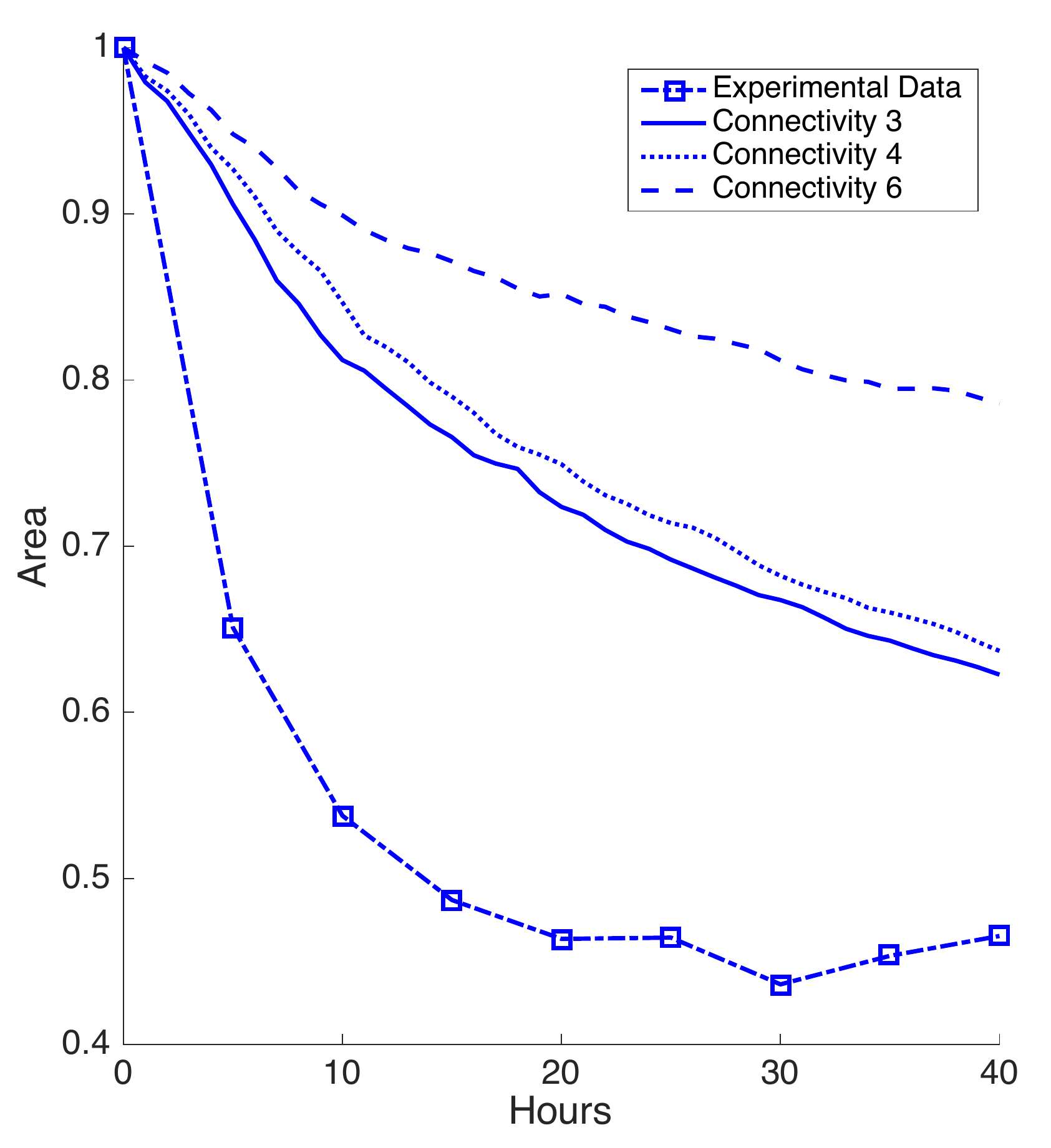}}

\caption{A comparison of contraction for cell densities of 30,000 cells per mL on the left and 100,000 cells per mL on the right when the collagen is represented by true springs.  The solid lines represent a minimum of 3 connections, the dotted lines represent a minimum of 4 connections and the dashed lines represent a minimum of 6 connections.  The squares denote experimental data.}
\label{fig:SpringDensity}
\end{figure}

Both the overall contraction rate as well as the final contraction amount is dependent on the interconnectivity of the collagen lattice.
Physically, the greater the number of connections that exist between elements in the collagen lattice, the greater the stiffness of the resulting lattice.  Several preliminary calculations were performed to determine approximately the optimal minimal connectivity of the collagen lattice.  In these calculations we were interested in determining a connectivity that would approximate the shape of the graph for each of the four experimental cell densities.  In Figure~\ref{fig:colcomp}, the simulated contraction is plotted against the experimental data.  Observe that when the collagen lattice is not stiff (due to low connectivity), those simulations involving high cell densities contract more than they should, and when the collagen lattice is too stiff (that is, the minimal number of connections is too high), the lattice does not contract enough at lower cell densities.  In our experiments in silico, we found a minimal connectivity between nodes of 30 connections to be ideal.  This is an order of magnitude higher than that found in \cite{Licup:2015:SCM,Sharma:2016:SCG} and 5 times higher than that predicted by Maxwell's criteria for rigidity \cite{Wyart:2008:EFS}.  In \cite{Notbohm:2015:MFP} a two-dimensional model with  microbuckling (the type of spring-like forces used here) was essential to produce correct force propogation with connectivity between 2 and 8.  But for three-dimensional models with microbuckling, they found that a connectivity of 14 gave reasonable results.  We therefore checked our model without microbuckling (i.e., true springs which resist compression).  The results are shown in Figure~\ref{fig:SpringDensity}.  All the parameters are the same except there is no microbuckling, and the connectivity of the lattices are 3, 4, and 6.  It is clear that the lattices are more stiff with lower connectivity than the model with microbuckling.  Thus a much higher connectivity should be used when microbuckling is assumed.

\subsection{Force propagation distances}
In \cite{Notbohm:2015:MFP} the authors concluded that microbuckling was essential to match experimental data regarding force propagation and to see tethers between cells.  Experimental data indicated that force decreased with a rate proportional to $r^{n}$ where $r$ is the distance from the cell and $n=-.52$.  In a three-dimensional model, they found connectivities of 3.5 and 14 gave results $n=-.82$ and $n=-.67$ respectively.  In their paper, the three-dimensional lattice was simulated with a regular grid of nodes connected in a cubic type pattern. Our model, in contrast, models a three-dimensional lattice with nodes in an irregular pattern and connections are randomly chosen as previously explained.  For our lattice, we calculated $n=-.858$ and $n=-.267$  with minimum connectivities of 6 and 30 respectively.  As the connectivity of our lattice increased, the distance the force propagated also increased as was seen in \cite{Notbohm:2015:MFP} in the presence of microbuckling. Our lattices with a minimum of 30 connections per node propagated force greater distances than in the experimental data.  This could be in part due to the sparse number of nodes in our model when compared to the volume of the lattice.  In order for the cell attachment sites to attach to lattice nodes in an approximately radially symmetric manner, the attachment sites had to be on the order of 100 microns away from the cell center.  The results were obtained by fitting data from 10 different lattices with nonlinear least squares.

\subsection{Direction of cell motion}\label{sec:isiteloc}
\begin{figure}
\begin{center}
\tikzset{cross/.style={cross out, draw, 
         minimum size=2*(#1-\pgflinewidth), 
         inner sep=0pt, outer sep=0pt},
cross/.default={1pt}}

\tdplotsetmaincoords{70}{110}
\begin{tikzpicture}[scale=.7,tdplot_main_coords]

\draw[color=black,thick,->] (0,0,0) -- (4,0,0) node[anchor=north east]{$x$};
\draw[color=black,thick,->] (0,0,0) -- (0,4,0) node[anchor=north west]{$y$};
\draw[color=black,thick,->] (0,0,0) -- (0,0,4) node[anchor=south]{$z$};
\shadedraw[left color=gray!50,right color=gray!50,draw=gray!50]
(0,0,0)--(4,4,4) --(0,1.05,3.8)--cycle;
\shadedraw[left color=gray!50,right color=gray!50,draw=gray!50]
(0,0,0)--(4,4,4) --(0,4.,1.1)--cycle;
\fill[
  top color=gray!80,
  bottom color=gray!80,
  rotate around={-45:(4,4,4)},
  ] 
  (3.97,3.97,3.97) circle (2cm and 0.5cm);
\draw[rotate around={15.8:(0,0,0)}]
  (0,0,0) -- (0,4.05,0) -- cycle;
 \node (origin) at (0,0,0){};
\draw[shift={(-3.89,-3.89,-1.55)},rotate around={89:(0,4+2,0)},rotate around={46.3:(origin)}]
 (0,4,0) arc (180:360:2cm and 0.5cm) ;

\draw[rotate around={81.6:(0,0,0)}]
  (0,0,0) -- (0,3.73,0) -- cycle;
 \node (origin) at (0,0,0){};
\draw[dashed,shift={(-3.89,-3.89,-1.55)},rotate around={88:(0,4+2,0)},rotate around={46.3:(origin)}]
 (0,4,0) arc (180:0:2cm and 0.5cm) ;

 \tikzstyle{spring}=[decorate,decoration={
        coil,
        amplitude = 3pt,
        aspect=.5,
        pre length=10pt,
        post length=10pt,
       segment length=3pt,
    }
]
\draw[line width=2pt, arrows = {-Stealth[length=12pt, inset=3pt]}]
(0,0,0) -- (4,4,4);
\tikzset{xyplane/.estyle={cm={
cos(\theta),sin(\theta)*sin(\psi),sin(\phi)*sin(\theta),
cos(\phi)*cos(\psi)-sin(\phi)*cos(\theta)*sin(\psi),#1},decorate,decoration={
        coil,
        amplitude = 3pt,
        aspect=.5,
        pre length=10pt,
        post length=10pt,
       segment length=3pt,
    }}}

\draw (0,0,0) node[circle,fill,inner sep=5pt](center){};;
\draw (3,2,3) node[line width=2pt,cross=6pt,color=black!45] (at1){};
\draw (-2,-5^.5,3) node[line width=2pt,cross=6pt] (at2){};
\draw (-2,5^.5,3) node[line width=2pt,cross=6pt] (at3){};
\draw (0,-3,0) node[line width=2pt,cross=6pt] (at4){};
\draw (2,-2,0) node[line width=2pt,cross=6pt] (at5){};
\draw (-3,2,0) node[line width=2pt,cross=6pt,color=black!65] (at6){};
\draw (-1,-2,0) node[line width=2pt,cross=6pt] (at7){};
\draw (0,3,-3) node[line width=2pt,cross=6pt] (at8){};
\draw (4,1,-3) node[line width=2pt,cross=6pt] (at9){};
\draw (4,5,-3) node[line width=2pt,cross=6pt] (at10){};
\draw[spring,color=black!45] (center) -- (at1);
\draw[spring](center)--(at2);
\draw[spring](center)--(at3);
\draw[spring](center)--(at4);
\draw[spring](center)--(at5);
\draw[spring,color=black!65](center)--(at6);
\draw[spring](center)--(at7);
\draw[spring](center)--(at8);
\draw[spring](center)--(at9);
\draw[spring](center)--(at10);
\end{tikzpicture}
\end{center}
\caption{The location of a new I-site depends on the direction of motion of the cell.}
\label{fig:dircell}
\end{figure}
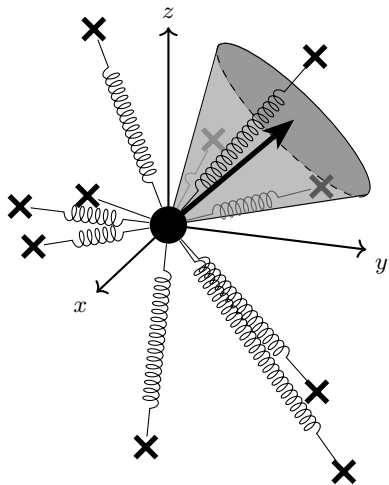

Prior to determining the appropriate value for the connectivity of the lattice, an important change to the placement of the I-sites occurred as compared to previous models by the authors.  In the current model, the direction of placement of the I-site is determined by the velocity of the cell.  In prior models, the placement of the I-site was dependent on the prior location of the I-site.  Specifically in the current model, when an integrin detaches from the collagen lattice, the current velocity of the cell is determined.  The I-site is then located with probability 0.8 in a cone with an opening angle of 4 degrees in the direction of the cell velocity, see Figure~\ref{fig:dircell}. (Initially, each cell is given a random velocity.)  This results in directed cell motion.  The introduction of the placement of the I-sites in the direction of motion allowed the model to reproduce two important features seen in biological experiments, namely, fibroblast distribution at the end of the simulation and the behavior of the collagen disc in the presence of a radial cut. Moreover adding directed cell motion eliminated the need for the cells to become inactive.  In previous work \cite{Dallon:2014:MMC}, in order to match the contraction of lattices in time and with different cell densities it was necessary to have the cell become inactive.  If the cells remained active they would continue contracting the lattice well beyond what was seen experimentally.  Thus it was postulated that a mechanosensing mechanism inactivated the cells.  This is no longer assumed. 

\subsubsection{Cell distribution at the conclusion of the simulation} 
\begin{figure}
\centerline{\includegraphics*[width=.34\linewidth]{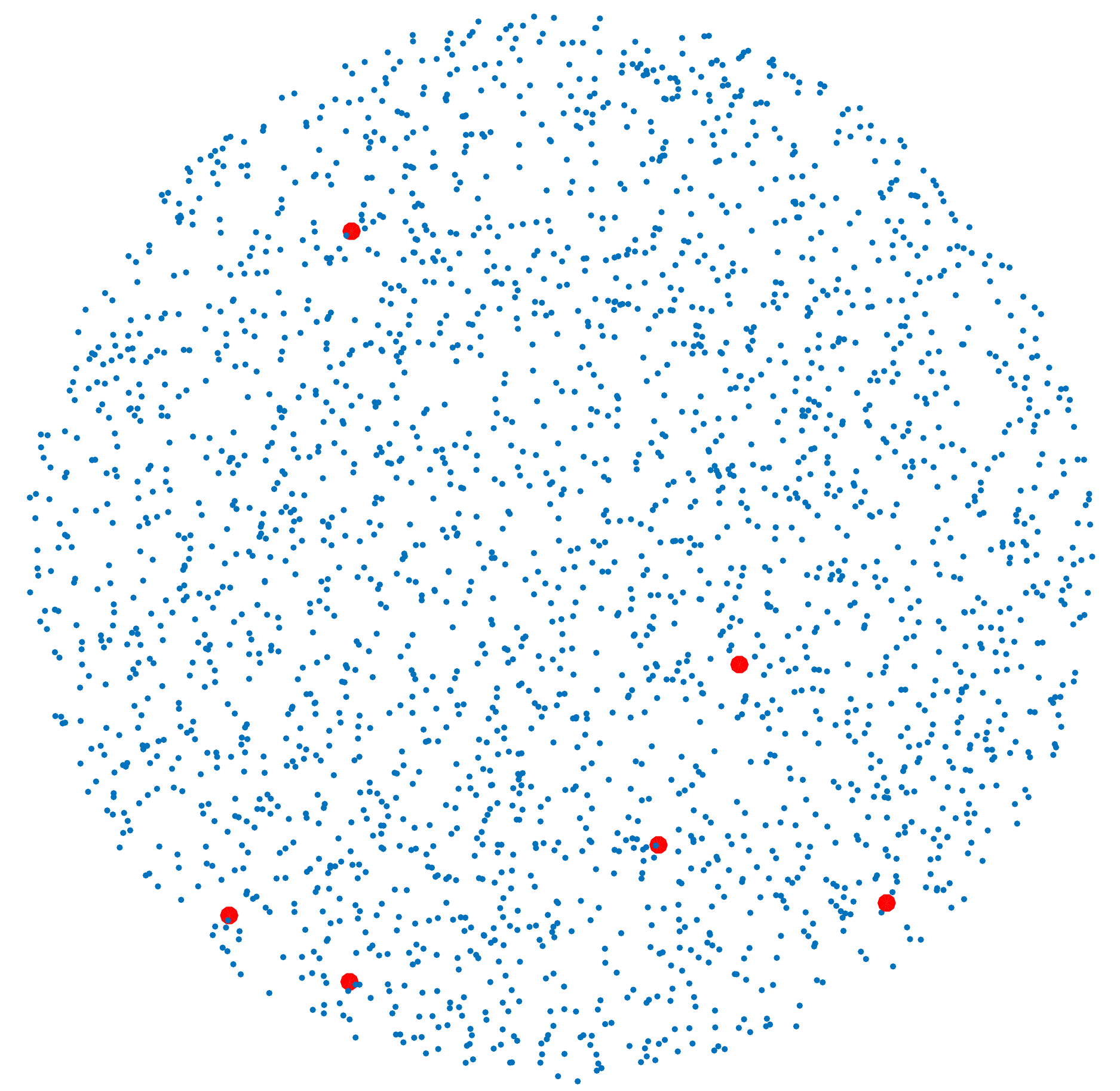}\quad\quad\includegraphics*[width=.34\linewidth]{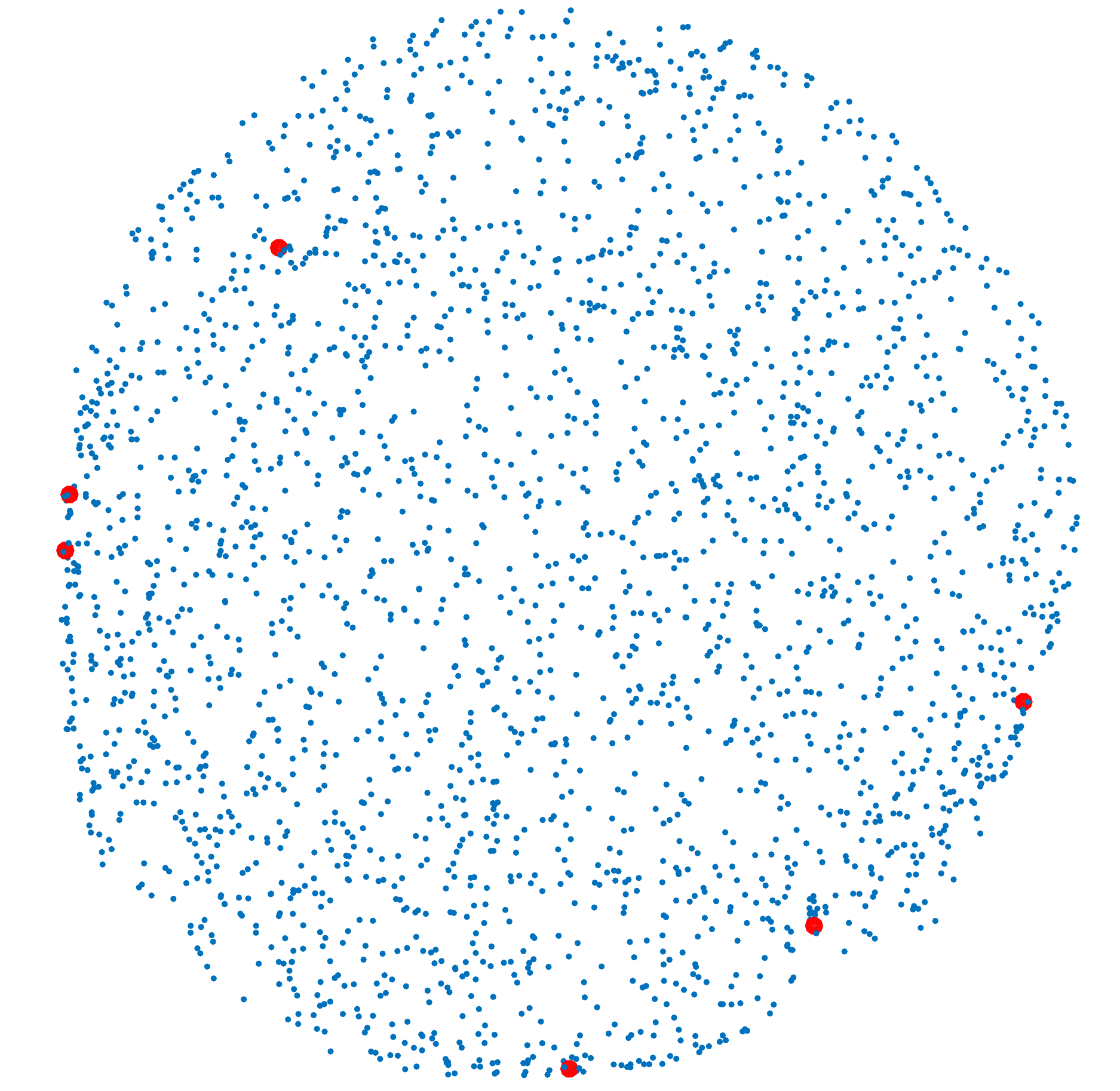}}\vspace{.2in}

\centerline{\includegraphics*[width=.34\linewidth]{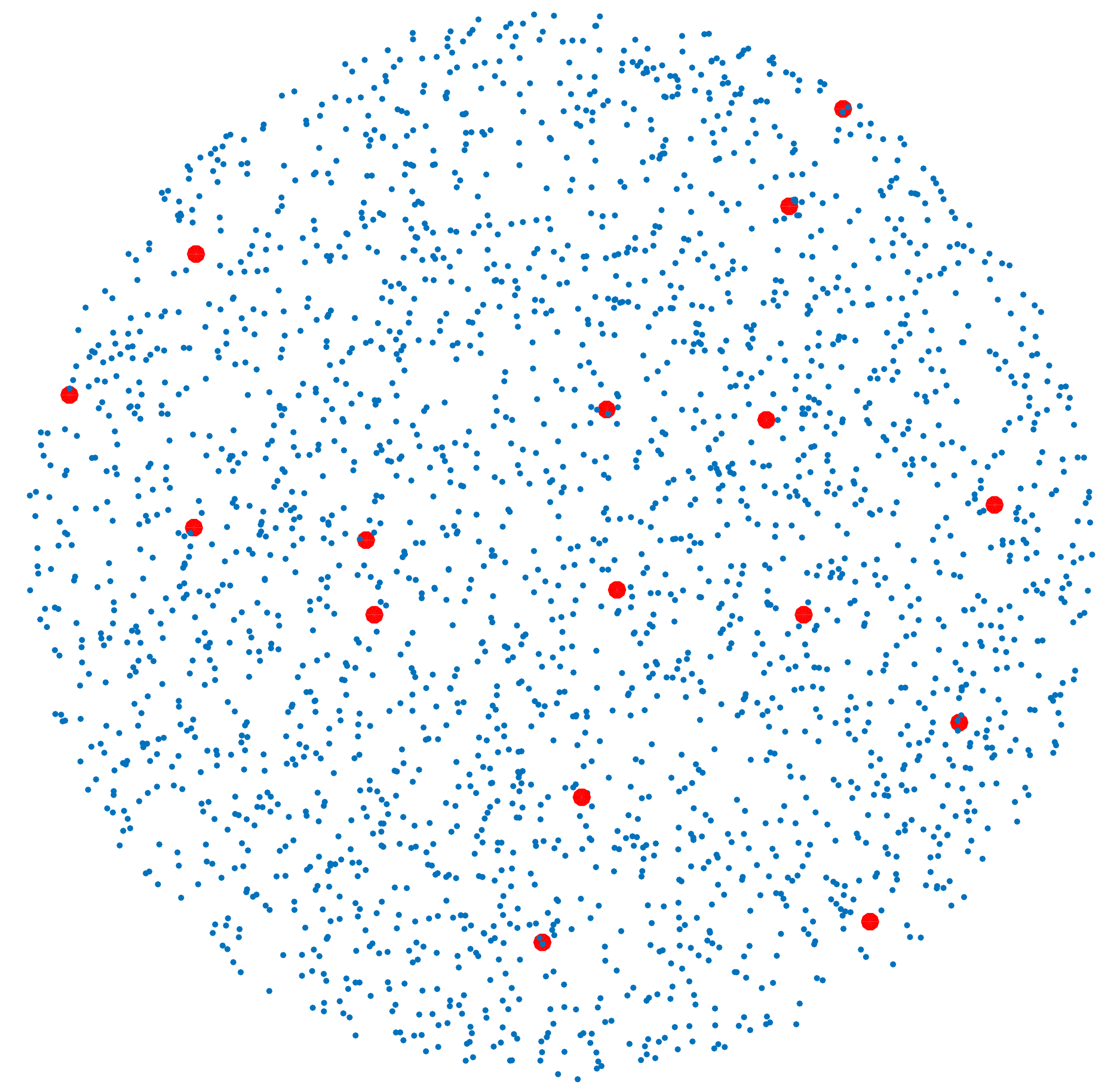}\quad\quad\includegraphics*[width=.34\linewidth]{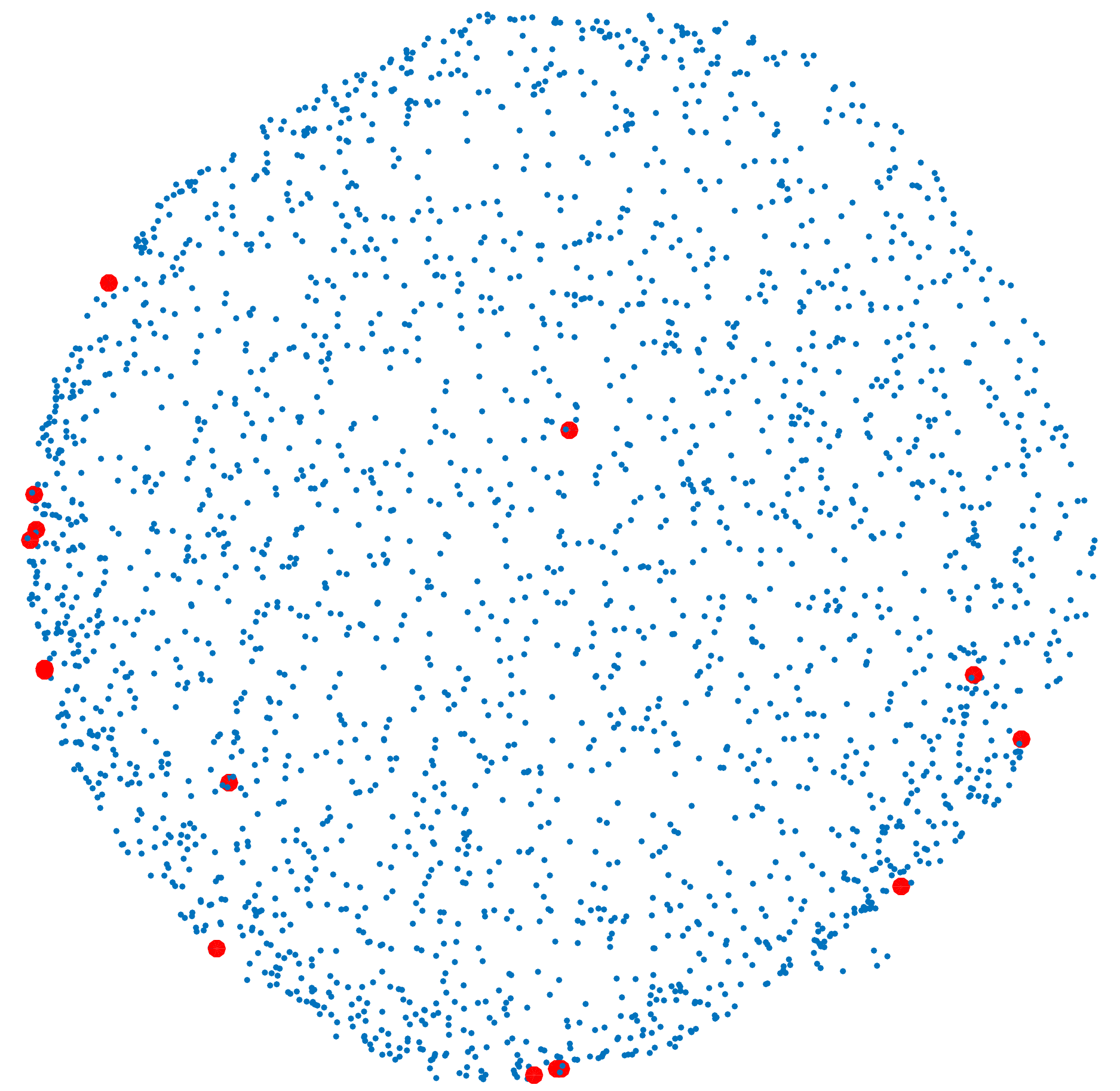}}\vspace{.2in}

\centerline{\includegraphics*[width=.34\linewidth]{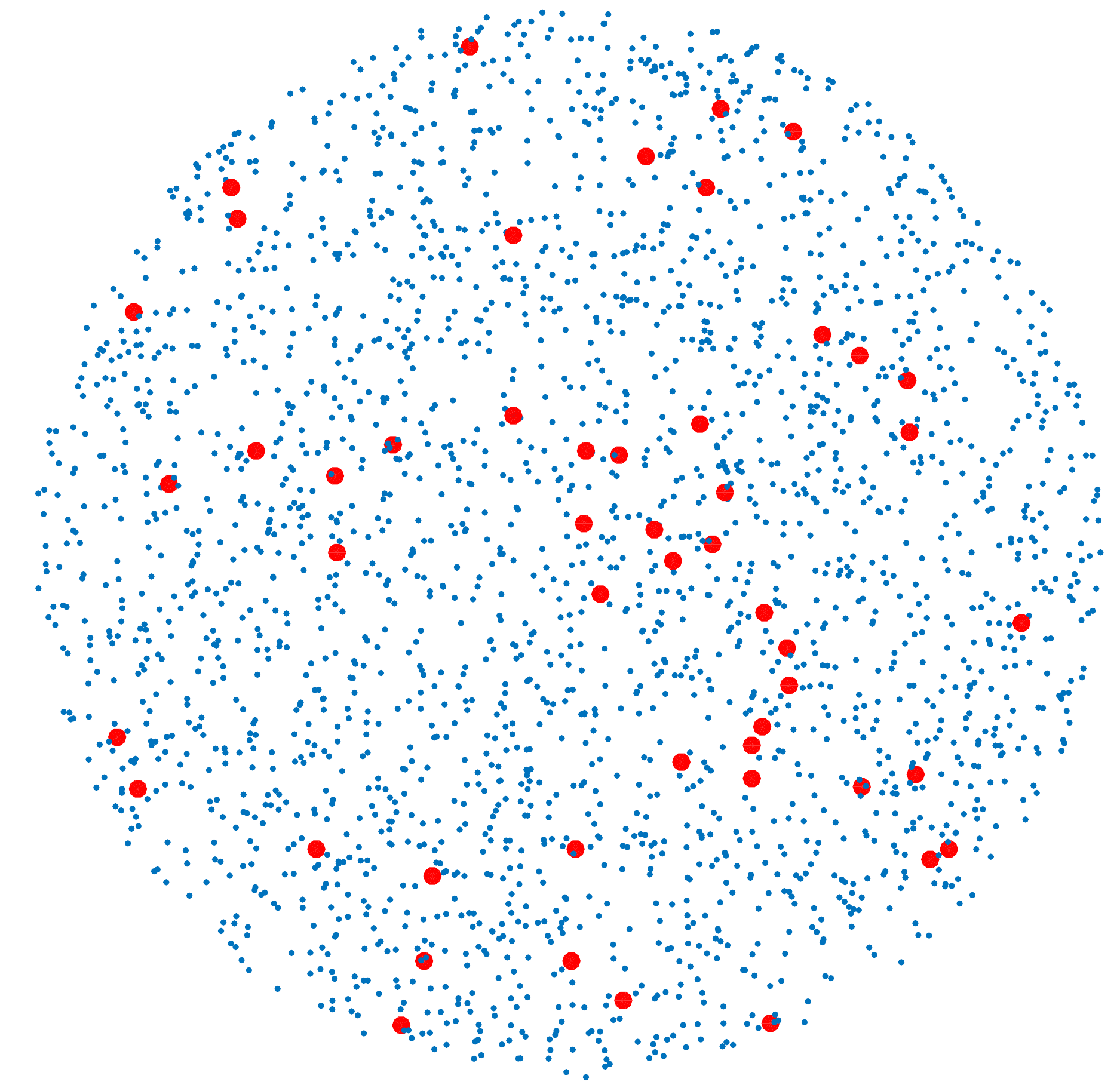}\quad\quad\includegraphics*[width=.34\linewidth]{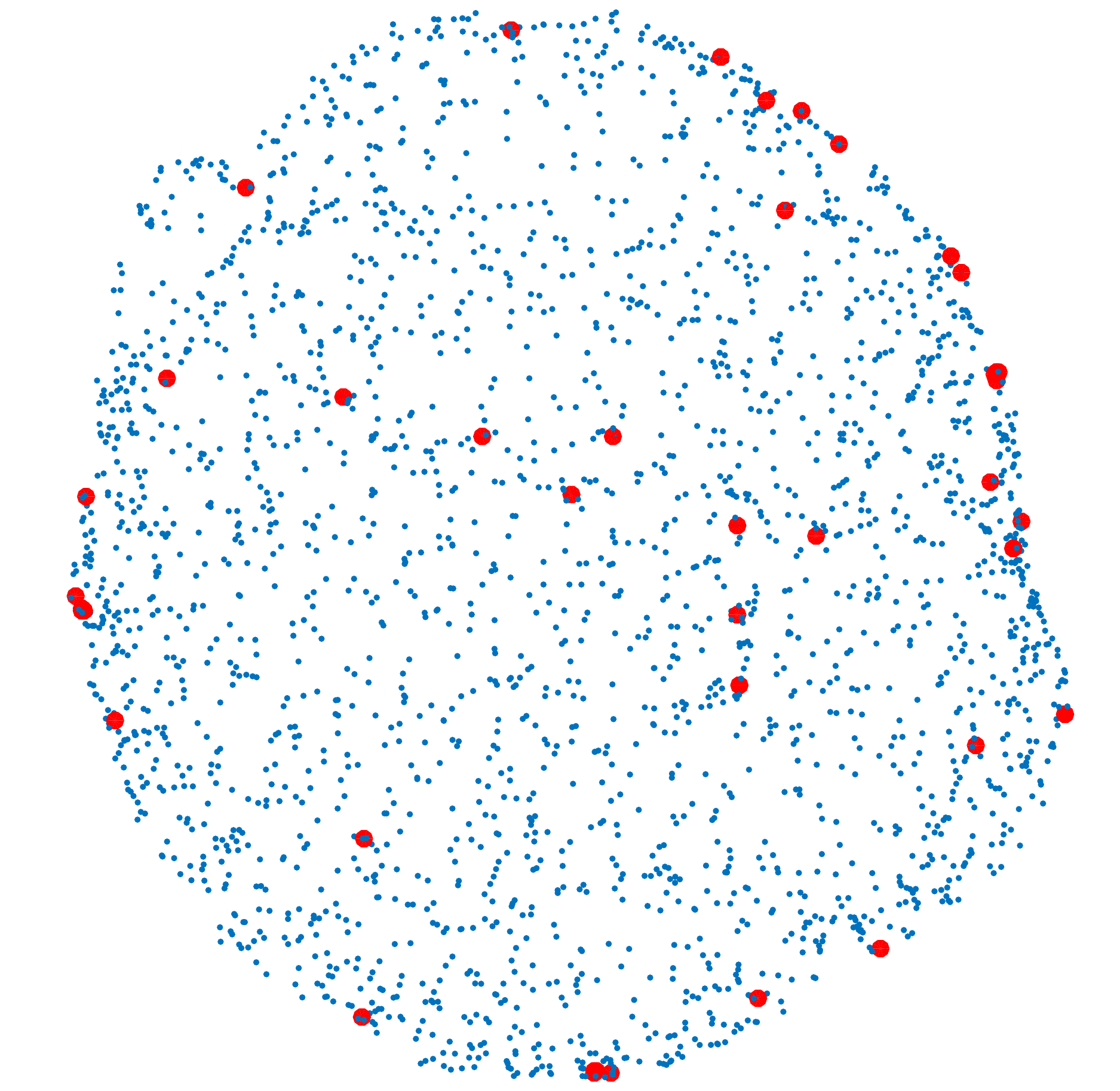}}\vspace{.2in}

\centerline{\includegraphics*[width=.34\linewidth]{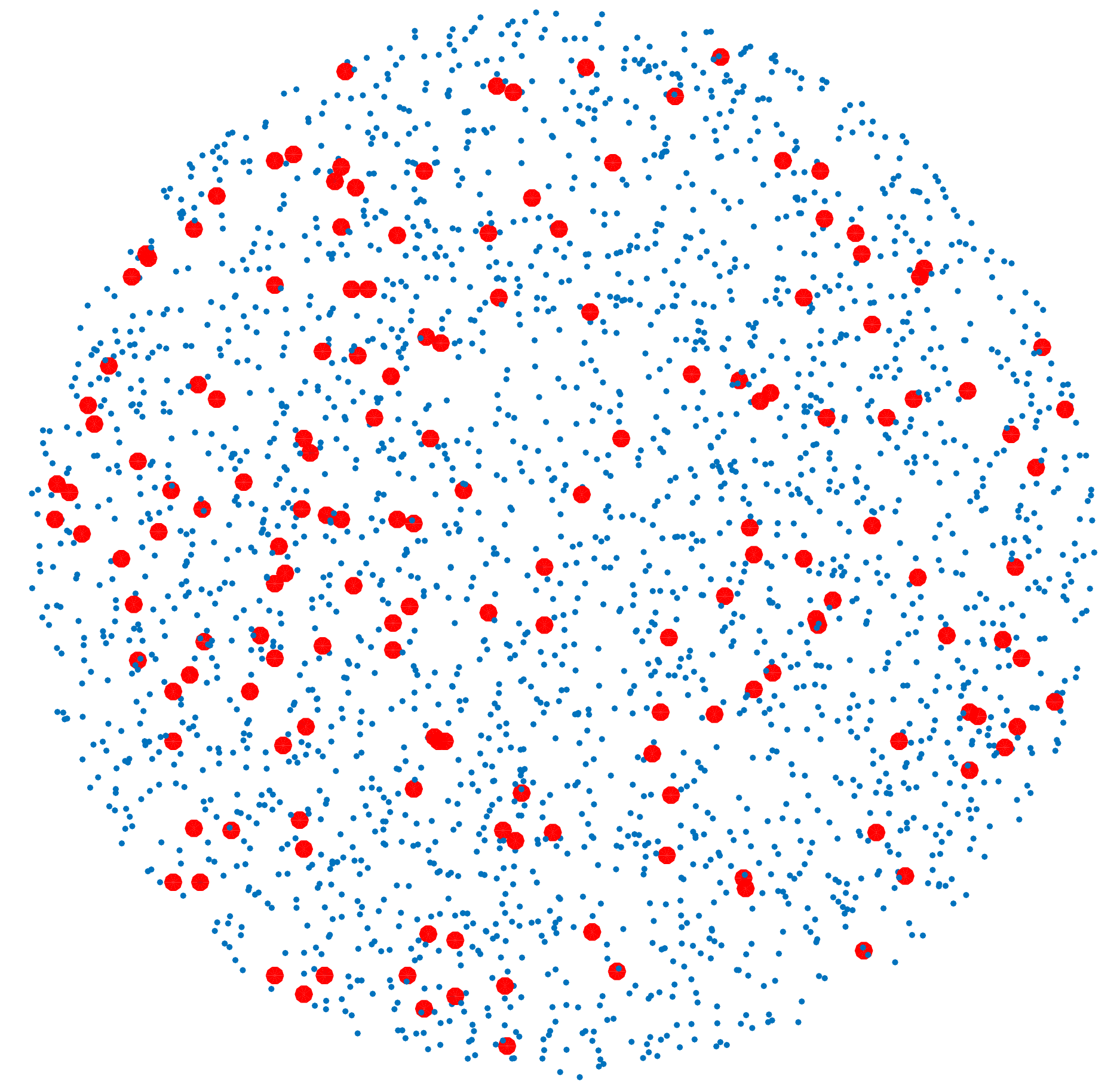}\quad\quad\includegraphics*[width=.34\linewidth]{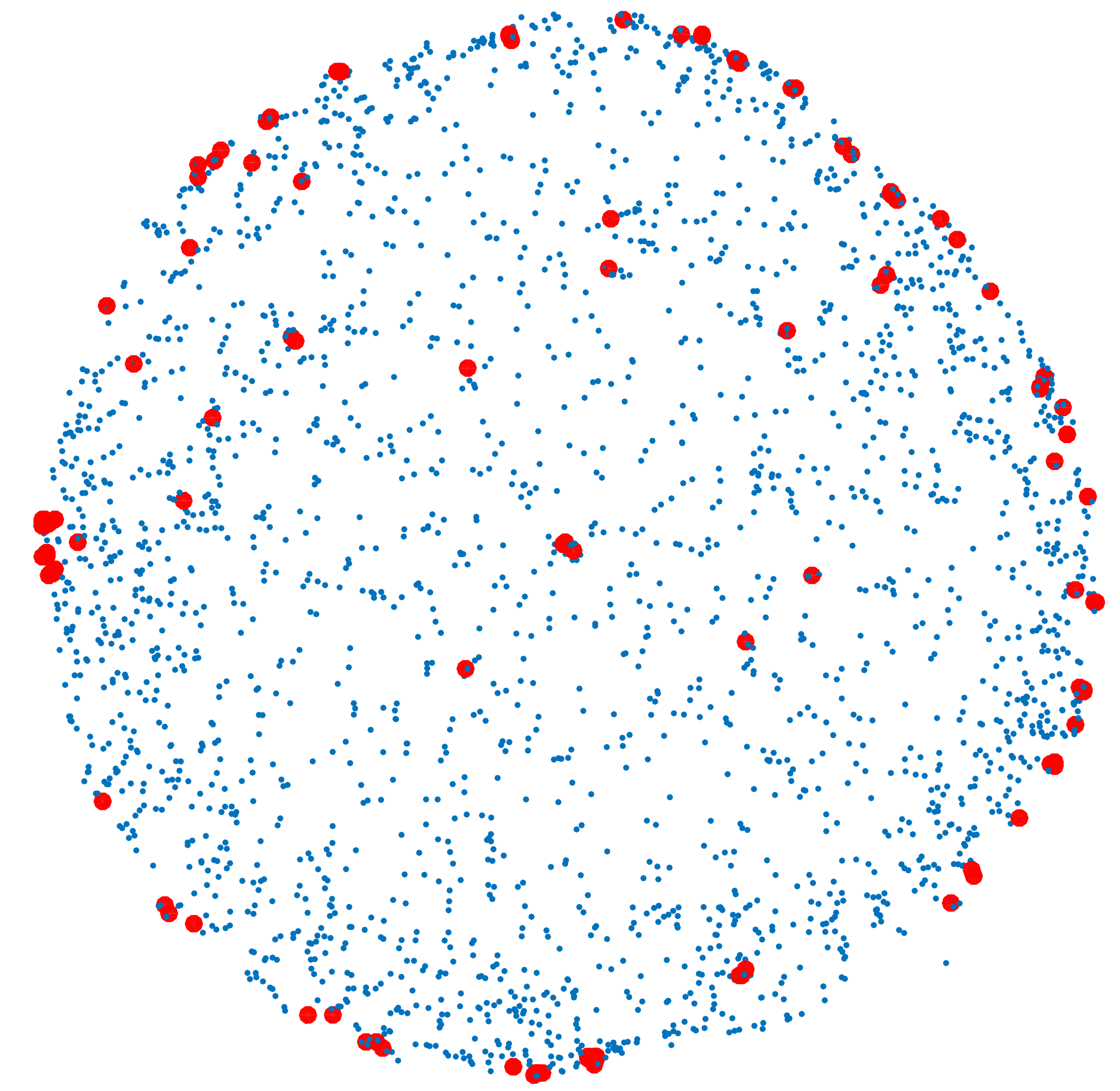}}
\caption{The initial and final configuration of the lattices are shown for four different densities with the cell center marked as solid circles.  The top row is the 3750 cells per
mL, the second row 10,000 cells per mL, the third row is 30,000 cells per mL, and the fourth row is
100,000 cells per mL.  On the left, the initial configuration is shown and on the right, the configuration is shown after 40 hours. }
\label{fig:firstlast}
\end{figure}

Experimentalists~\cite{Simon:2012:MRB,Ehrlich:2000:DMH} have observed that as cells contract a collagen lattice, the cells are more concentrated near the boundary of the lattice and
the lattice is also more compacted around the edge.  Figure~\ref{fig:firstlast}
shows simulation results for the initial cell distribution on the left and final cell distribution on the right for each
cell density.  It is easily seen that the cells are more dense near the boundary at the end of the simulations.

\begin{figure}
\centerline{\includegraphics*[width=.34\linewidth]{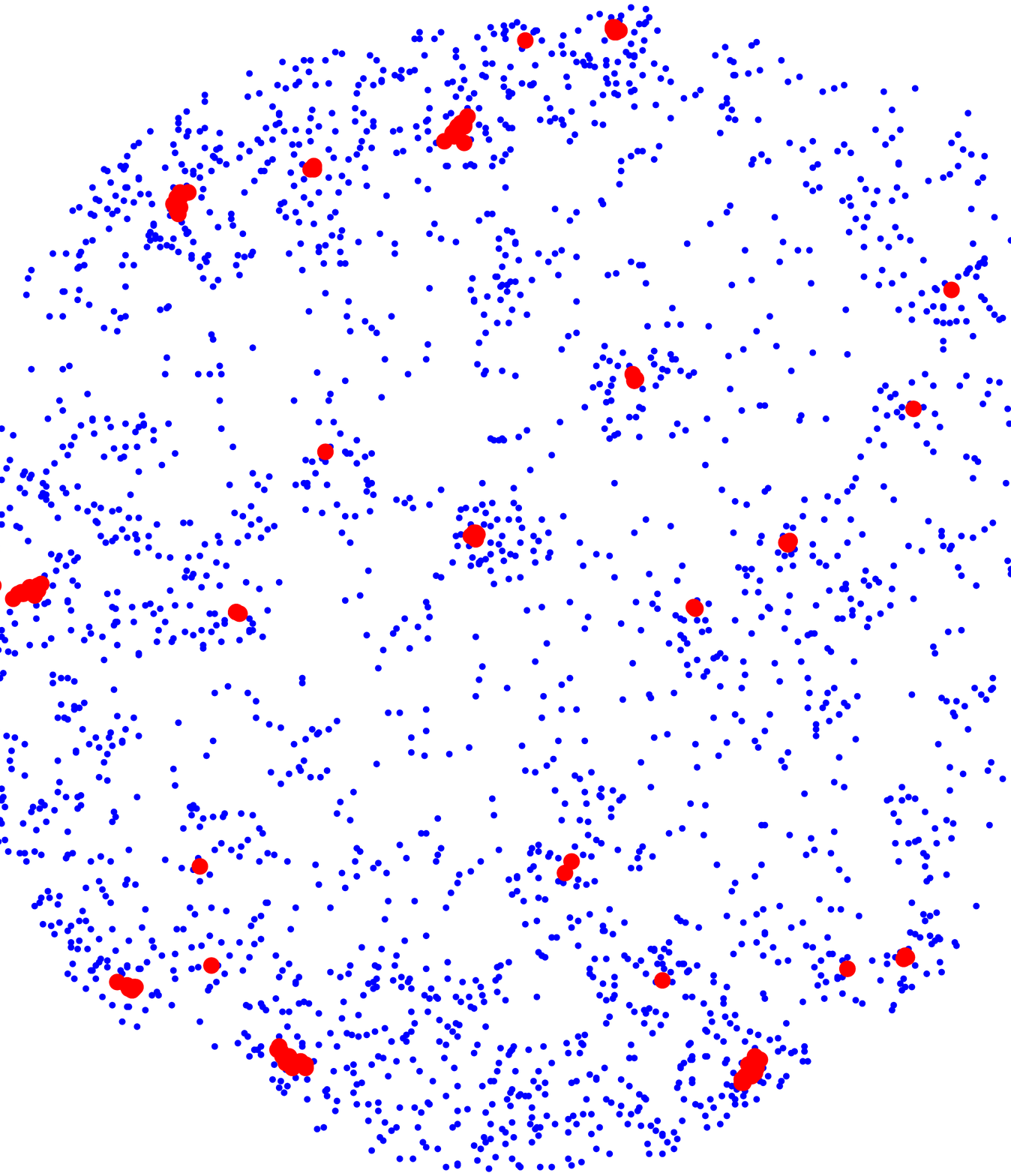}\quad\quad\quad\includegraphics*[width=.34\linewidth]{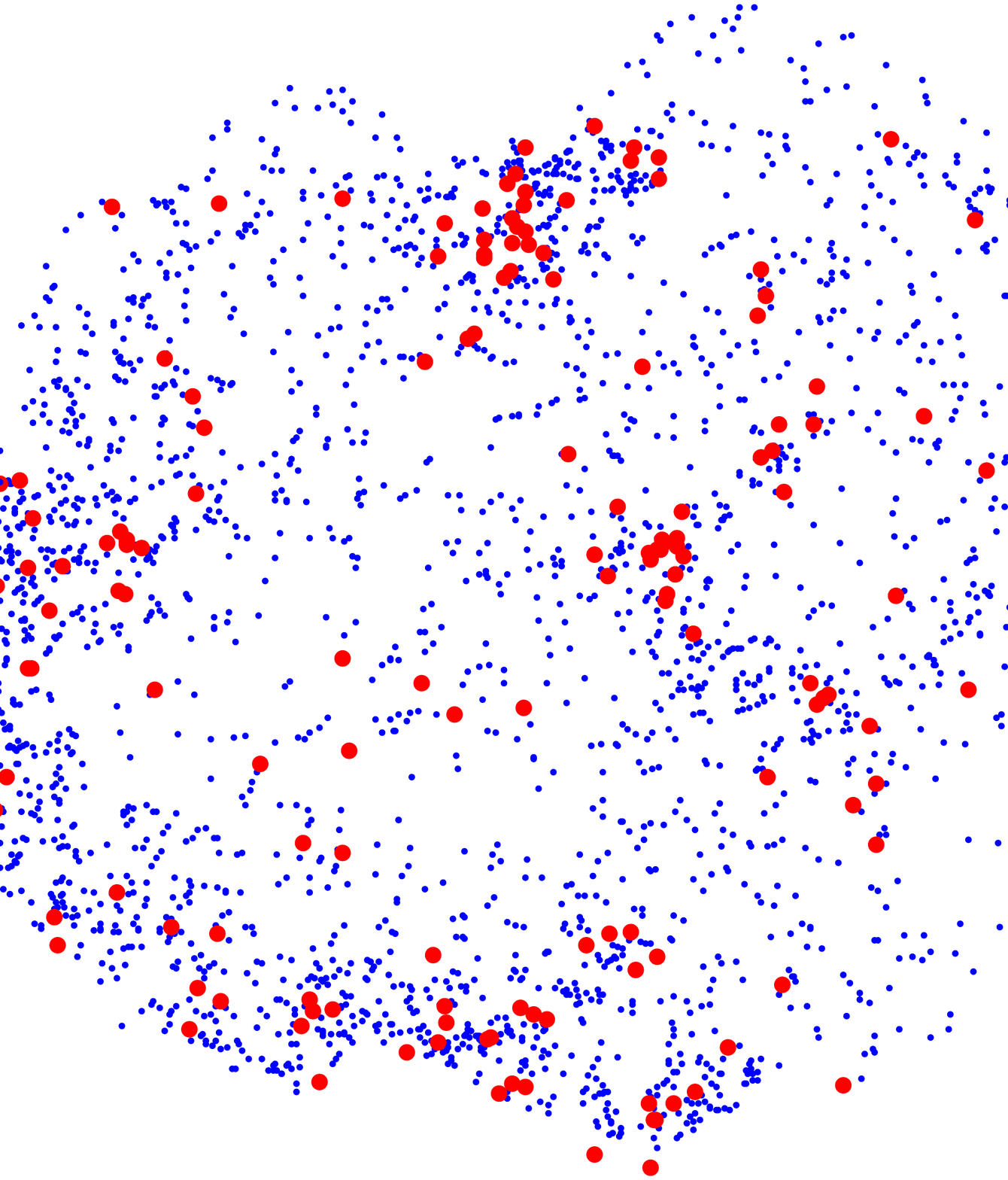}}
\caption{The final configuration after 40 hours of the lattices are shown for 100,000 cells per mL with the cell center marked as solid circles.  On the left is the final configuration in the case of undirected integrin placement and on the right the final configuration is shown when the cells become inactive due to mechanosensing. }
\label{fig:comparefinaldist}
\end{figure}

\begin{figure}
\centerline{\includegraphics[width= .8\linewidth]{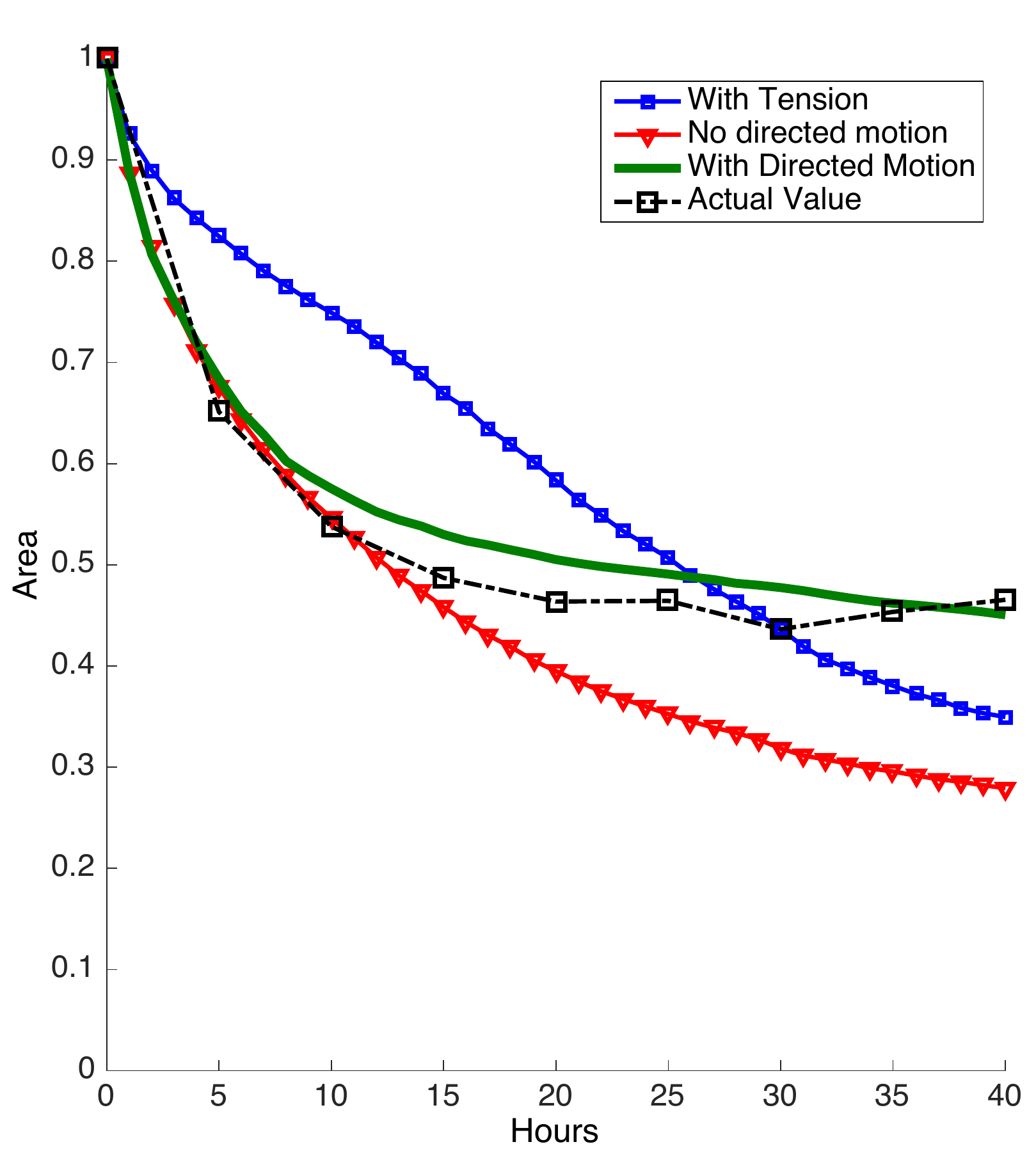}}
\caption{The lattice contraction for 100,000 cells per mL with directed motion, undirected motion, and mechanosensing rules (with tension)
compared to the experimental data.}
\label{fig:comprop}
\end{figure}

In Figure~\ref{fig:comparefinaldist} we showcase the effect of undirected integrin placement (on the left) and the mechanosensing rules of the prior paper which inactivate the cells (on the right) in the three-dimensional setting.  In both cases, the cells do not appear to be aggregating near the periphery of the lattice unlike the two-dimensional simulations \cite{Dallon:2014:MMC}.  Additionally, as the lattices contract they do not maintain a nice circular shape as is commonly seen experimentally and in the two-dimensional simulations.   In Figure~\ref{fig:comprop} we compare the collagen contraction for the mechanosensing rules to the new rule of directed motion.  Again the directed cell motion gives results which more closely match the experimental data.

In order to quantify the aggregation of cells near the periphery, we compare the interior cell density and the boundary cell density at the conclusion of the simulation with 100,000 cells per mL.  In particular, we consider the interior of the domain to be the 90\% of the domain closest to the center, and the boundary to be the remaining 10\% of the domain.  We then counted the number of cells in the interior and the number of cells at the boundary for 10 simulation runs with the same
initial lattice configuration but different random initial positions
for the cells and different instantiations for the random variables for the directed cell motion.  The results confirm the same characteristic of the
final cell distribution with the cell distribution at the boundary over 14 times more dense than the
density in the interior of the domain.  

\subsubsection{Radial cuts}
\begin{figure}
\begin{center}\includegraphics[width= .5\linewidth]{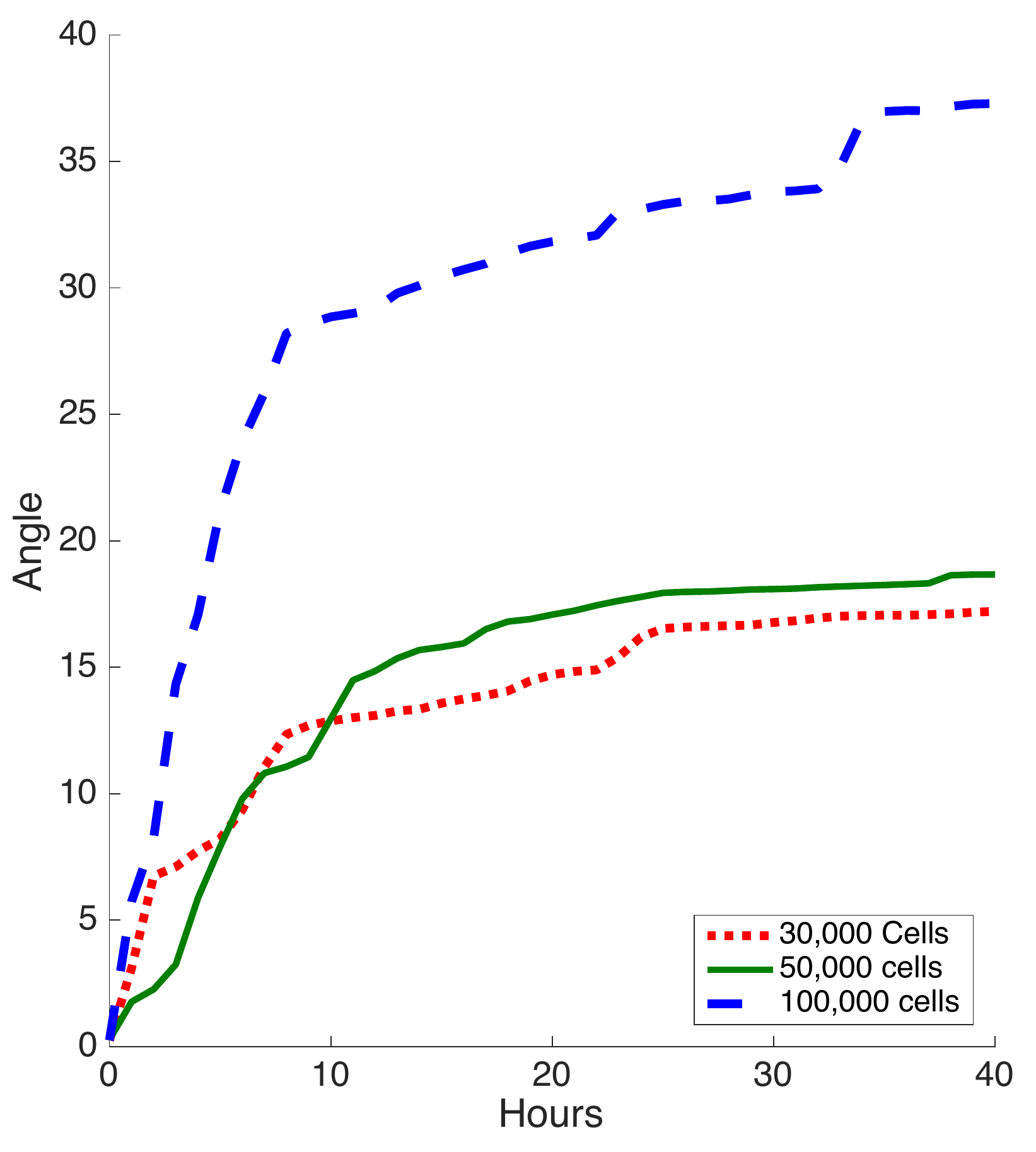}
\caption{A comparison of opening angles for cell densities of 30,000, 50,000, and 100,000 per mL.  Observe that for the 50,000 per mL the ending angle is approximately 16 degrees which corresponds to the results in \cite{Simon:2012:MRB}.}
\label{fig:angleplots}
\end{center}
\end{figure}

Experimentally~\cite{Simon:2012:MRB} radial cuts are used to determine the presence of residual type stress.  When a radial cut is introduced at the beginning of an experiment, only a small angle is initially seen (presumably due to the width of the blade).  When cuts are introduced one to five days into the experiment, the observed behavior is that a ``Pac-Man"-like shape forms where the angle of the opening is on average about $20\pm 6$ degrees~\cite{Simon:2012:MRB}.  We replicate this experiment by introducing a radial cut  and measuring the opening angle as the simulation progresses over 40 hours.  Assuming the center of the disc to be located at the origin, we determine the opening angle by calculating, for each collagen node with positive $x$ value, the angle determined by $\arctan(y/x)$.  We then classify these calculated angles into two categories, those with negative angle values (corresponding to negative $y$ values) and those with positive angle values (corresponding to the nodes with positive $y$ values). Our opening angle is then defined to be $\min(\text{positive angles}) - \max(\text{negative angles})$. In Figure~\ref{fig:angleplots} the opening angle, as a function of time, is given for differing cell densities.  For cell density of 50,000 cells per mL, 10 simulation runs with the same
initial lattice configuration but different random initial positions
for the cells and different instantiations for the random variables for the directed cell motion were performed and the average ending angle was 16 degrees, corresponding nicely with the experimental data.

\begin{figure}
\begin{center}\includegraphics[width= .4\linewidth]{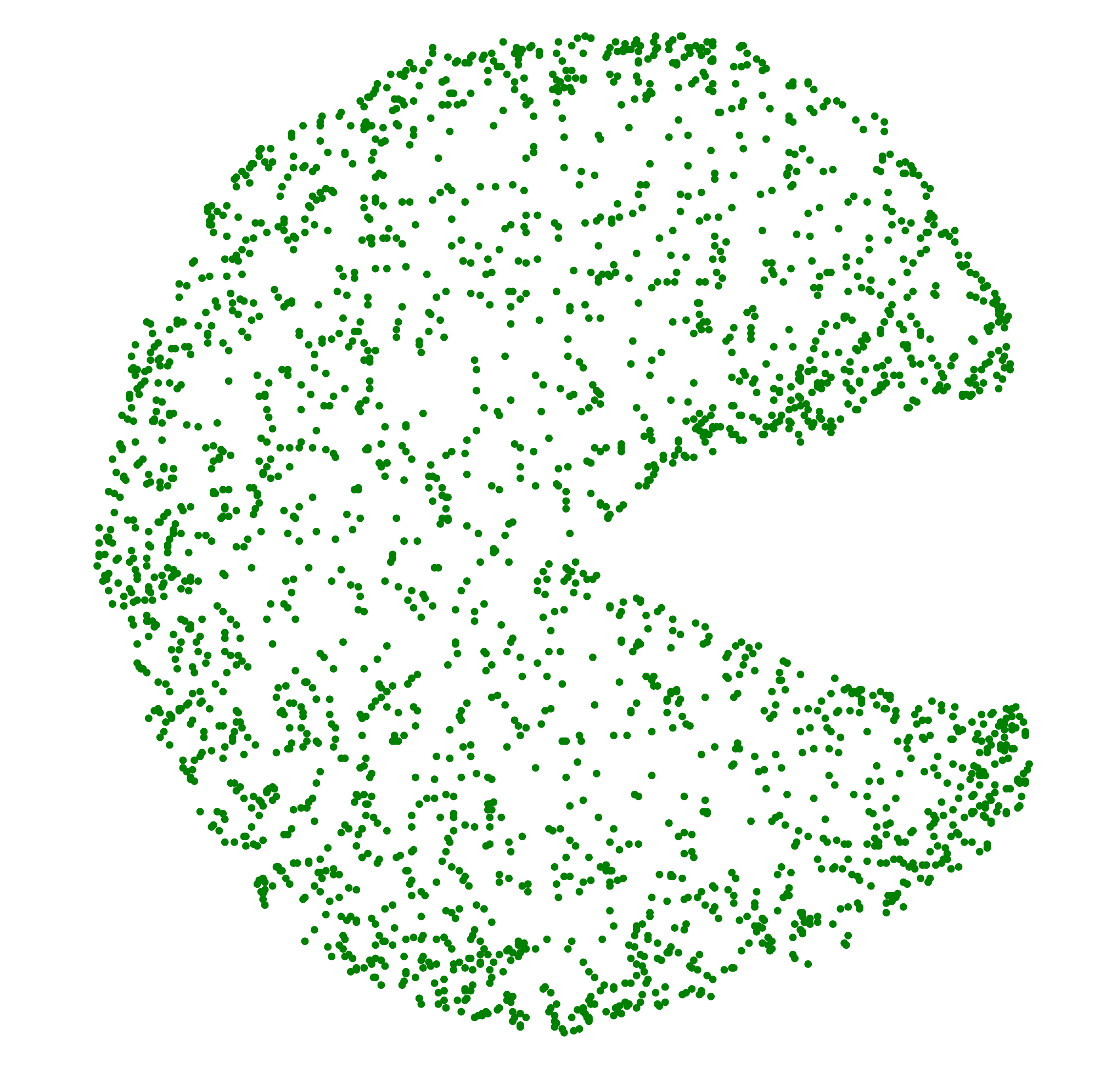}\hspace{.2in}\includegraphics[width= .4\linewidth]{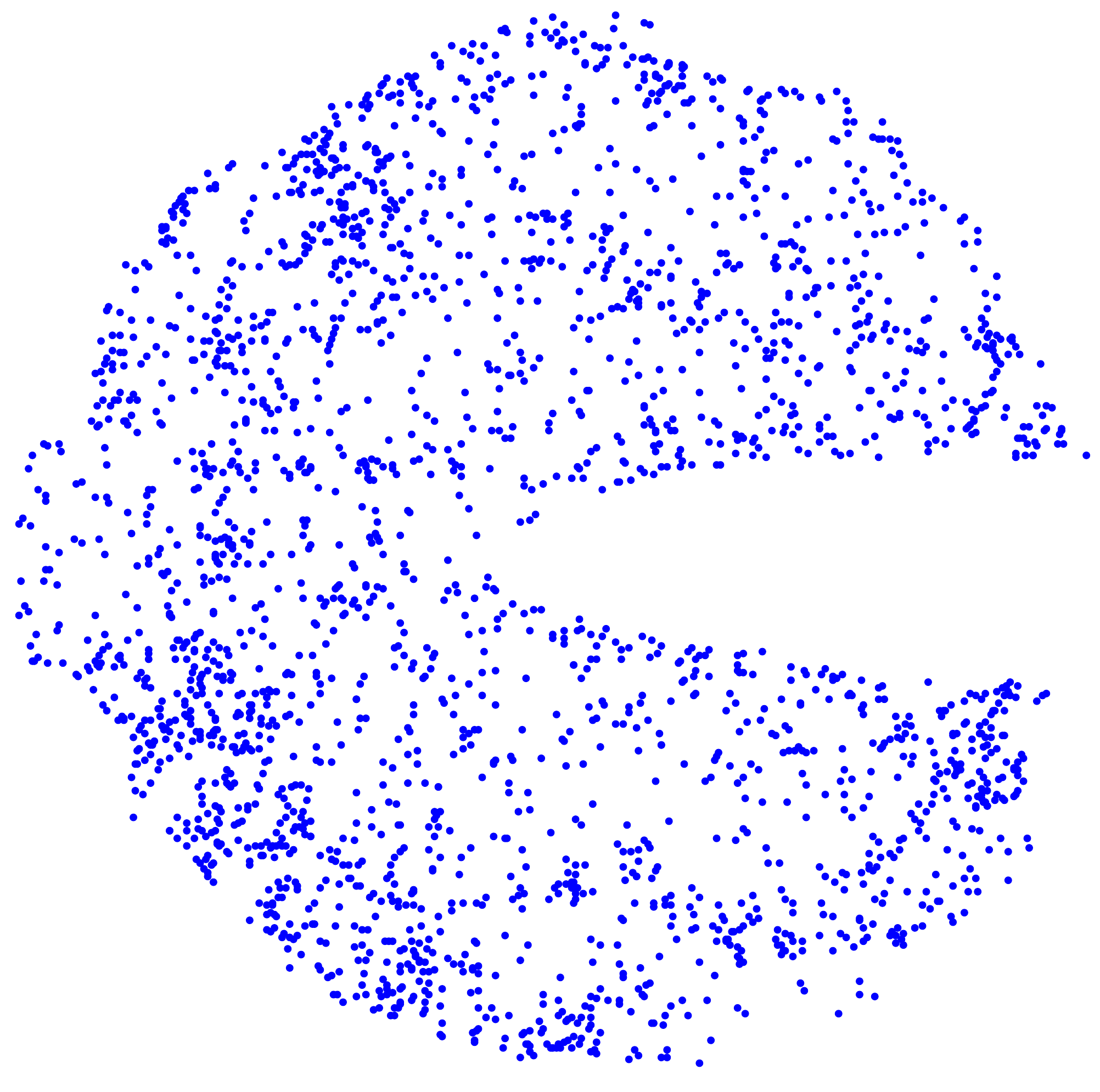}
\end{center}
\caption{A comparison of directed (on the left) vs. non-directed (on the right) motion with connection density of 30 and cell density of 100,000 cells per mL.  Observe that in the directed case the disc maintains it's circular shape, and does not develop teeth-like protrusions.}
\label{fig:teeth}
\end{figure}

Directed cell motion was not only necessary to obtain the approximately correct angle opening measurements, but also necessary to preserve the expected topology of the disc with a radial cut.  When the cell motion is undirected, the disc does not maintain the expected circular shape, but instead develops a bulge opposite the radial cut.  Moreover, the radial opening develops teeth-like protrusions along the boundary, see Figure~\ref{fig:teeth}.

\section{Discussion}
We have shown that when changing from a two-dimensional model to a three-dimensional model of a fibrous structure like collagen, the construct of the model lattice is crucial to model results.  For the two-dimensional model, a Delaunay triangulation worked well, but in order to match contraction data over a period of time for gels with several different cell densities, the model cells needed to become inactive or they would continue contracting the gels beyond what is seen experimentally \cite{Dallon:2014:MMC}.  When moving to a three-dimensional model, the analog of the two-dimensional Delaunay triangulation gave strange topology.  We changed the topology of the lattice and found that in order to match the previously mentioned data, with almost the same parameters for the collagen properties as the two-dimensional model, the connectivity was much higher than that predicted by theory and other discrete models of collagen \cite{Wyart:2008:EFS,Sharma:2016:SCG,Licup:2015:SCM} which did not include microbuckling.  When modeling with spring-like connections which do not resist compression, the connectivity must be much higher in order to match experimental data. To the best of our knowledge, all the connectivity values for collagen are determined by modeling and there is no experimental data for connectivity of collagen lattices.

We also found that with the new formulation of the lattice in three dimensions, the cells do not need to become inactive as in the two-dimensional gels if they exhibit directed motion. In other words, the cells need to be engaged in directed cell motion (the direction does not matter) by placing new attachment sites in the direction of motion.  With this assumption, the model results matched the final cell distribution data, the angle in the slit gel experiments, the contraction data, force-distance propagation data, and the overall morphology of the gels.

In conclusion, we have developed a three-dimensional model which, with reasonable assumptions, replicates the results of several different biological experiments.  The model assumes microbuckling, directed cell motion, and the same collagen properties as a similar two-dimensional model.  When modeling a three-dimensional lattice with microbuckling, the connectivity results of models with true spring interactions do not apply.  Assuming that cells are engaged in directed cell motion gives better results than undirected cell or force-sensing cells which inactivate.  Finally, connectivity and lattice generation greatly affect model results and need more systematic investigation.

\bibliography{lattice}

\begin{thebibliography}{21}%
\makeatletter
\providecommand \@ifxundefined [1]{%
 \@ifx{#1\undefined}
}%
\providecommand \@ifnum [1]{%
 \ifnum #1\expandafter \@firstoftwo
 \else \expandafter \@secondoftwo
 \fi
}%
\providecommand \@ifx [1]{%
 \ifx #1\expandafter \@firstoftwo
 \else \expandafter \@secondoftwo
 \fi
}%
\providecommand \natexlab [1]{#1}%
\providecommand \enquote  [1]{``#1''}%
\providecommand \bibnamefont  [1]{#1}%
\providecommand \bibfnamefont [1]{#1}%
\providecommand \citenamefont [1]{#1}%
\providecommand \href@noop [0]{\@secondoftwo}%
\providecommand \href [0]{\begingroup \@sanitize@url \@href}%
\providecommand \@href[1]{\@@startlink{#1}\@@href}%
\providecommand \@@href[1]{\endgroup#1\@@endlink}%
\providecommand \@sanitize@url [0]{\catcode `\\12\catcode `\$12\catcode
  `\&12\catcode `\#12\catcode `\^12\catcode `\_12\catcode `\%12\relax}%
\providecommand \@@startlink[1]{}%
\providecommand \@@endlink[0]{}%
\providecommand \url  [0]{\begingroup\@sanitize@url \@url }%
\providecommand \@url [1]{\endgroup\@href {#1}{\urlprefix }}%
\providecommand \urlprefix  [0]{URL }%
\providecommand \Eprint [0]{\href }%
\providecommand \doibase [0]{http://dx.doi.org/}%
\providecommand \selectlanguage [0]{\@gobble}%
\providecommand \bibinfo  [0]{\@secondoftwo}%
\providecommand \bibfield  [0]{\@secondoftwo}%
\providecommand \translation [1]{[#1]}%
\providecommand \BibitemOpen [0]{}%
\providecommand \bibitemStop [0]{}%
\providecommand \bibitemNoStop [0]{.\EOS\space}%
\providecommand \EOS [0]{\spacefactor3000\relax}%
\providecommand \BibitemShut  [1]{\csname bibitem#1\endcsname}%
\let\auto@bib@innerbib\@empty
\bibitem [{\citenamefont {Dallon}\ \emph {et~al.}(2014)\citenamefont {Dallon},
  \citenamefont {Evans},\ and\ \citenamefont {Ehrlich}}]{Dallon:2014:MMC}%
  \BibitemOpen
  \bibfield  {author} {\bibinfo {author} {\bibfnamefont {J.~C.}\ \bibnamefont
  {Dallon}}, \bibinfo {author} {\bibfnamefont {E.~J.}\ \bibnamefont {Evans}}, \
  and\ \bibinfo {author} {\bibfnamefont {H.}~\bibnamefont {Ehrlich}},\ }\href
  {\doibase 10.1098/rsif.2014.0598} {\bibfield  {journal} {\bibinfo  {journal}
  {Journal of the Royal Society Interface}\ }\textbf {\bibinfo {volume} {11}}
  (\bibinfo {year} {2014}),\ 10.1098/rsif.2014.0598}\BibitemShut {NoStop}%
\bibitem [{\citenamefont {Bell}\ \emph {et~al.}(1979)\citenamefont {Bell},
  \citenamefont {Ivarsson},\ and\ \citenamefont {Charlotte}}]{Bell:1979:PTS}%
  \BibitemOpen
  \bibfield  {author} {\bibinfo {author} {\bibfnamefont {E.}~\bibnamefont
  {Bell}}, \bibinfo {author} {\bibfnamefont {B.}~\bibnamefont {Ivarsson}}, \
  and\ \bibinfo {author} {\bibfnamefont {M.}~\bibnamefont {Charlotte}},\
  }\href@noop {} {\bibfield  {journal} {\bibinfo  {journal} {Proce. Natl. Acad.
  Sci.}\ }\textbf {\bibinfo {volume} {76}},\ \bibinfo {pages} {1274} (\bibinfo
  {year} {1979})}\BibitemShut {NoStop}%
\bibitem [{\citenamefont {Ehrlich}\ \emph {et~al.}(2006)\citenamefont
  {Ehrlich}, \citenamefont {Sun}, \citenamefont {Kainth},\ and\ \citenamefont
  {Kromah}}]{Ehrlich:2006:EMW}%
  \BibitemOpen
  \bibfield  {author} {\bibinfo {author} {\bibfnamefont {H.~P.}\ \bibnamefont
  {Ehrlich}}, \bibinfo {author} {\bibfnamefont {B.}~\bibnamefont {Sun}},
  \bibinfo {author} {\bibfnamefont {K.~S.}\ \bibnamefont {Kainth}}, \ and\
  \bibinfo {author} {\bibfnamefont {F.}~\bibnamefont {Kromah}},\ }\href@noop {}
  {\bibfield  {journal} {\bibinfo  {journal} {Wound Repair and Regeneration}\
  }\textbf {\bibinfo {volume} {14}},\ \bibinfo {pages} {625} (\bibinfo {year}
  {2006})}\BibitemShut {NoStop}%
\bibitem [{\citenamefont {Greco}\ and\ \citenamefont
  {Ehrlich}(1992)}]{Greco:1992:DCD}%
  \BibitemOpen
  \bibfield  {author} {\bibinfo {author} {\bibfnamefont {R.~M.}\ \bibnamefont
  {Greco}}\ and\ \bibinfo {author} {\bibfnamefont {H.~P.}\ \bibnamefont
  {Ehrlich}},\ }\href@noop {} {\bibfield  {journal} {\bibinfo  {journal}
  {Tissue and Cell}\ }\textbf {\bibinfo {volume} {24}},\ \bibinfo {pages} {843}
  (\bibinfo {year} {1992})}\BibitemShut {NoStop}%
\bibitem [{\citenamefont {Martin-Martin}\ \emph {et~al.}(2011)\citenamefont
  {Martin-Martin}, \citenamefont {Tovell}, \citenamefont {Dahlmann-Noor},
  \citenamefont {Khaw},\ and\ \citenamefont {Bailly}}]{Martin:2011:EMI}%
  \BibitemOpen
  \bibfield  {author} {\bibinfo {author} {\bibfnamefont {B.}~\bibnamefont
  {Martin-Martin}}, \bibinfo {author} {\bibfnamefont {V.}~\bibnamefont
  {Tovell}}, \bibinfo {author} {\bibfnamefont {A.~H.}\ \bibnamefont
  {Dahlmann-Noor}}, \bibinfo {author} {\bibfnamefont {P.~T.}\ \bibnamefont
  {Khaw}}, \ and\ \bibinfo {author} {\bibfnamefont {M.}~\bibnamefont
  {Bailly}},\ }\href@noop {} {\bibfield  {journal} {\bibinfo  {journal}
  {European journal of cell biology}\ }\textbf {\bibinfo {volume} {90}},\
  \bibinfo {pages} {26} (\bibinfo {year} {2011})}\BibitemShut {NoStop}%
\bibitem [{\citenamefont {Redden}\ and\ \citenamefont
  {Doolin}(2003)}]{Redden:2003:CCC}%
  \BibitemOpen
  \bibfield  {author} {\bibinfo {author} {\bibfnamefont {R.~A.}\ \bibnamefont
  {Redden}}\ and\ \bibinfo {author} {\bibfnamefont {E.~J.}\ \bibnamefont
  {Doolin}},\ }\href@noop {} {\bibfield  {journal} {\bibinfo  {journal} {Skin
  Research and Technology}\ }\textbf {\bibinfo {volume} {9}},\ \bibinfo {pages}
  {290} (\bibinfo {year} {2003})}\BibitemShut {NoStop}%
\bibitem [{\citenamefont {Dallon}\ and\ \citenamefont
  {Ehrlich}(2008)}]{Dallon:2008:RFC}%
  \BibitemOpen
  \bibfield  {author} {\bibinfo {author} {\bibfnamefont {J.~C.}\ \bibnamefont
  {Dallon}}\ and\ \bibinfo {author} {\bibfnamefont {H.~P.}\ \bibnamefont
  {Ehrlich}},\ }\href {\doibase 10.1111/j.1524-475X.2008.00392.x} {\bibfield
  {journal} {\bibinfo  {journal} {Wound Repair Regen}\ }\textbf {\bibinfo
  {volume} {16}},\ \bibinfo {pages} {472} (\bibinfo {year} {2008})}\BibitemShut
  {NoStop}%
\bibitem [{\citenamefont {Wyart}\ \emph {et~al.}(2008)\citenamefont {Wyart},
  \citenamefont {Liang}, \citenamefont {Kabla},\ and\ \citenamefont
  {Mahadevan}}]{Wyart:2008:EFS}%
  \BibitemOpen
  \bibfield  {author} {\bibinfo {author} {\bibfnamefont {M.}~\bibnamefont
  {Wyart}}, \bibinfo {author} {\bibfnamefont {H.}~\bibnamefont {Liang}},
  \bibinfo {author} {\bibfnamefont {A.}~\bibnamefont {Kabla}}, \ and\ \bibinfo
  {author} {\bibfnamefont {L.}~\bibnamefont {Mahadevan}},\ }\href@noop {}
  {\bibfield  {journal} {\bibinfo  {journal} {Physical review letters}\
  }\textbf {\bibinfo {volume} {101}},\ \bibinfo {pages} {215501} (\bibinfo
  {year} {2008})}\BibitemShut {NoStop}%
\bibitem [{\citenamefont {Sharma}\ \emph {et~al.}(2016)\citenamefont {Sharma},
  \citenamefont {Licup}, \citenamefont {Jansen}, \citenamefont {Rens},
  \citenamefont {Sheinman}, \citenamefont {Koenderink},\ and\ \citenamefont
  {MacKintosh}}]{Sharma:2016:SCG}%
  \BibitemOpen
  \bibfield  {author} {\bibinfo {author} {\bibfnamefont {A.}~\bibnamefont
  {Sharma}}, \bibinfo {author} {\bibfnamefont {A.~J.}\ \bibnamefont {Licup}},
  \bibinfo {author} {\bibfnamefont {K.~A.}\ \bibnamefont {Jansen}}, \bibinfo
  {author} {\bibfnamefont {R.}~\bibnamefont {Rens}}, \bibinfo {author}
  {\bibfnamefont {M.}~\bibnamefont {Sheinman}}, \bibinfo {author}
  {\bibfnamefont {G.~H.}\ \bibnamefont {Koenderink}}, \ and\ \bibinfo {author}
  {\bibfnamefont {F.~C.}\ \bibnamefont {MacKintosh}},\ }\href
  {http://dx.doi.org/10.1038/nphys3628} {\bibfield  {journal} {\bibinfo
  {journal} {Nat Phys}\ }\textbf {\bibinfo {volume} {12}},\ \bibinfo {pages}
  {584} (\bibinfo {year} {2016})}\BibitemShut {NoStop}%
\bibitem [{\citenamefont {Licup}\ \emph {et~al.}(2015)\citenamefont {Licup},
  \citenamefont {M{\"u}nster}, \citenamefont {Sharma}, \citenamefont
  {Sheinman}, \citenamefont {Jawerth}, \citenamefont {Fabry}, \citenamefont
  {Weitz},\ and\ \citenamefont {MacKintosh}}]{Licup:2015:SCM}%
  \BibitemOpen
  \bibfield  {author} {\bibinfo {author} {\bibfnamefont {A.~J.}\ \bibnamefont
  {Licup}}, \bibinfo {author} {\bibfnamefont {S.}~\bibnamefont {M{\"u}nster}},
  \bibinfo {author} {\bibfnamefont {A.}~\bibnamefont {Sharma}}, \bibinfo
  {author} {\bibfnamefont {M.}~\bibnamefont {Sheinman}}, \bibinfo {author}
  {\bibfnamefont {L.~M.}\ \bibnamefont {Jawerth}}, \bibinfo {author}
  {\bibfnamefont {B.}~\bibnamefont {Fabry}}, \bibinfo {author} {\bibfnamefont
  {D.~A.}\ \bibnamefont {Weitz}}, \ and\ \bibinfo {author} {\bibfnamefont
  {F.~C.}\ \bibnamefont {MacKintosh}},\ }\href {\doibase
  10.1073/pnas.1504258112} {\bibfield  {journal} {\bibinfo  {journal} {Proc
  Natl Acad Sci U S A}\ }\textbf {\bibinfo {volume} {112}},\ \bibinfo {pages}
  {9573} (\bibinfo {year} {2015})}\BibitemShut {NoStop}%
\bibitem [{\citenamefont {Barocas}\ and\ \citenamefont
  {Tranquillo}(1997)}]{Barocas:1997:ABT}%
  \BibitemOpen
  \bibfield  {author} {\bibinfo {author} {\bibfnamefont {V.~H.}\ \bibnamefont
  {Barocas}}\ and\ \bibinfo {author} {\bibfnamefont {R.~T.}\ \bibnamefont
  {Tranquillo}},\ }\href@noop {} {\ \textbf {\bibinfo {volume} {119}},\
  \bibinfo {pages} {137} (\bibinfo {year} {1997})}\BibitemShut {NoStop}%
\bibitem [{\citenamefont {Dallon}\ and\ \citenamefont
  {Othmer}(1998)}]{Dallon:1998:CAC}%
  \BibitemOpen
  \bibfield  {author} {\bibinfo {author} {\bibfnamefont {J.~C.}\ \bibnamefont
  {Dallon}}\ and\ \bibinfo {author} {\bibfnamefont {H.~G.}\ \bibnamefont
  {Othmer}},\ }\href@noop {} {\ \textbf {\bibinfo {volume} {194}},\ \bibinfo
  {pages} {461} (\bibinfo {year} {1998})}\BibitemShut {NoStop}%
\bibitem [{\citenamefont {Olsen}\ \emph {et~al.}(1999)\citenamefont {Olsen},
  \citenamefont {Maini}, \citenamefont {Sherratt},\ and\ \citenamefont
  {Dallon}}]{Olsen:1999:MMA}%
  \BibitemOpen
  \bibfield  {author} {\bibinfo {author} {\bibfnamefont {L.}~\bibnamefont
  {Olsen}}, \bibinfo {author} {\bibfnamefont {P.~K.}\ \bibnamefont {Maini}},
  \bibinfo {author} {\bibfnamefont {J.~A.}\ \bibnamefont {Sherratt}}, \ and\
  \bibinfo {author} {\bibfnamefont {J.~C.}\ \bibnamefont {Dallon}},\
  }\href@noop {} {\ \textbf {\bibinfo {volume} {158}},\ \bibinfo {pages} {145}
  (\bibinfo {year} {1999})}\BibitemShut {NoStop}%
\bibitem [{\citenamefont {Dallon}\ \emph {et~al.}(1999)\citenamefont {Dallon},
  \citenamefont {Sherratt},\ and\ \citenamefont {Maini}}]{Dallon:1999:MME}%
  \BibitemOpen
  \bibfield  {author} {\bibinfo {author} {\bibfnamefont {J.~C.}\ \bibnamefont
  {Dallon}}, \bibinfo {author} {\bibfnamefont {J.~A.}\ \bibnamefont
  {Sherratt}}, \ and\ \bibinfo {author} {\bibfnamefont {P.~K.}\ \bibnamefont
  {Maini}},\ }\href@noop {} {\ \textbf {\bibinfo {volume} {199}},\ \bibinfo
  {pages} {449} (\bibinfo {year} {1999})},\ \bibinfo {note} {{\tt
  http://www.academicpress.com/jtb}}\BibitemShut {NoStop}%
\bibitem [{\citenamefont {Schl{\"u}ter}\ \emph {et~al.}(2012)\citenamefont
  {Schl{\"u}ter}, \citenamefont {Ramis-Conde},\ and\ \citenamefont
  {Chaplain}}]{Schluter:2012:CMS}%
  \BibitemOpen
  \bibfield  {author} {\bibinfo {author} {\bibfnamefont {D.~K.}\ \bibnamefont
  {Schl{\"u}ter}}, \bibinfo {author} {\bibfnamefont {I.}~\bibnamefont
  {Ramis-Conde}}, \ and\ \bibinfo {author} {\bibfnamefont {M.~A.~J.}\
  \bibnamefont {Chaplain}},\ }\href {\doibase 10.1016/j.bpj.2012.07.048}
  {\bibfield  {journal} {\bibinfo  {journal} {Biophys J}\ }\textbf {\bibinfo
  {volume} {103}},\ \bibinfo {pages} {1141} (\bibinfo {year}
  {2012})}\BibitemShut {NoStop}%
\bibitem [{\citenamefont {Reinhardt}\ and\ \citenamefont
  {Gooch}(2014)}]{Reinhardt:2014:AMT}%
  \BibitemOpen
  \bibfield  {author} {\bibinfo {author} {\bibfnamefont {J.~W.}\ \bibnamefont
  {Reinhardt}}\ and\ \bibinfo {author} {\bibfnamefont {K.~J.}\ \bibnamefont
  {Gooch}},\ }\href {\doibase 10.1115/1.4026179} {\bibfield  {journal}
  {\bibinfo  {journal} {J Biomech Eng}\ }\textbf {\bibinfo {volume} {136}},\
  \bibinfo {pages} {021024} (\bibinfo {year} {2014})}\BibitemShut {NoStop}%
\bibitem [{\citenamefont {Notbohm}\ \emph {et~al.}(2015)\citenamefont
  {Notbohm}, \citenamefont {Lesman}, \citenamefont {Rosakis}, \citenamefont
  {Tirrell},\ and\ \citenamefont {Ravichandran}}]{Notbohm:2015:MFP}%
  \BibitemOpen
  \bibfield  {author} {\bibinfo {author} {\bibfnamefont {J.}~\bibnamefont
  {Notbohm}}, \bibinfo {author} {\bibfnamefont {A.}~\bibnamefont {Lesman}},
  \bibinfo {author} {\bibfnamefont {P.}~\bibnamefont {Rosakis}}, \bibinfo
  {author} {\bibfnamefont {D.~A.}\ \bibnamefont {Tirrell}}, \ and\ \bibinfo
  {author} {\bibfnamefont {G.}~\bibnamefont {Ravichandran}},\ }\href@noop {}
  {\bibfield  {journal} {\bibinfo  {journal} {Journal of The Royal Society
  Interface}\ }\textbf {\bibinfo {volume} {12}},\ \bibinfo {pages} {20150320}
  (\bibinfo {year} {2015})}\BibitemShut {NoStop}%
\bibitem [{\citenamefont {Simon}\ \emph {et~al.}(2012)\citenamefont {Simon},
  \citenamefont {Horgan},\ and\ \citenamefont {Humphrey}}]{Simon:2012:MRB}%
  \BibitemOpen
  \bibfield  {author} {\bibinfo {author} {\bibfnamefont {D.}~\bibnamefont
  {Simon}}, \bibinfo {author} {\bibfnamefont {C.}~\bibnamefont {Horgan}}, \
  and\ \bibinfo {author} {\bibfnamefont {J.}~\bibnamefont {Humphrey}},\
  }\href@noop {} {\bibfield  {journal} {\bibinfo  {journal} {Journal of the
  mechanical behavior of biomedical materials}\ }\textbf {\bibinfo {volume}
  {14}},\ \bibinfo {pages} {216} (\bibinfo {year} {2012})}\BibitemShut
  {NoStop}%
\bibitem [{\citenamefont {Simon}\ \emph {et~al.}(2014)\citenamefont {Simon},
  \citenamefont {Murtada},\ and\ \citenamefont {Humphrey}}]{Simon:2014:CMM}%
  \BibitemOpen
  \bibfield  {author} {\bibinfo {author} {\bibfnamefont {D.}~\bibnamefont
  {Simon}}, \bibinfo {author} {\bibfnamefont {S.-I.}\ \bibnamefont {Murtada}},
  \ and\ \bibinfo {author} {\bibfnamefont {J.}~\bibnamefont {Humphrey}},\
  }\href@noop {} {\bibfield  {journal} {\bibinfo  {journal} {International
  journal for numerical methods in biomedical engineering}\ }\textbf {\bibinfo
  {volume} {30}},\ \bibinfo {pages} {1506} (\bibinfo {year}
  {2014})}\BibitemShut {NoStop}%
\bibitem [{\citenamefont {Dallon}\ \emph {et~al.}(2013)\citenamefont {Dallon},
  \citenamefont {Scott},\ and\ \citenamefont {Smith}}]{Dallon:2013:FBM}%
  \BibitemOpen
  \bibfield  {author} {\bibinfo {author} {\bibfnamefont {J.~C.}\ \bibnamefont
  {Dallon}}, \bibinfo {author} {\bibfnamefont {M.}~\bibnamefont {Scott}}, \
  and\ \bibinfo {author} {\bibfnamefont {W.~V.}\ \bibnamefont {Smith}},\ }\href
  {\doibase 10.1115/1.4023987} {\bibfield  {journal} {\bibinfo  {journal}
  {Journal of Biomechanical Engineering}\ }\textbf {\bibinfo {volume} {135}},\
  \bibinfo {pages} {071008} (\bibinfo {year} {2013})}\BibitemShut {NoStop}%
\bibitem [{\citenamefont {Ehrlich}\ and\ \citenamefont
  {Rittenberg}(2000)}]{Ehrlich:2000:DMH}%
  \BibitemOpen
  \bibfield  {author} {\bibinfo {author} {\bibfnamefont {H.~P.}\ \bibnamefont
  {Ehrlich}}\ and\ \bibinfo {author} {\bibfnamefont {T.}~\bibnamefont
  {Rittenberg}},\ }\href@noop {} {\bibfield  {journal} {\bibinfo  {journal}
  {Journal of Cellular Physiology}\ }\textbf {\bibinfo {volume} {185}},\
  \bibinfo {pages} {432} (\bibinfo {year} {2000})}\BibitemShut {NoStop}%
\end{thebibliography}%

\end{document}